\documentclass[12pt]{spieman}  
\usepackage{amsmath,amsfonts,amssymb}
\usepackage{graphicx}
\usepackage{setspace}
\usepackage{tocloft}
\usepackage[version=4]{mhchem}
\usepackage[numbers, sort&compress]{natbib}
\usepackage{xcolor}

\usepackage{fancyhdr}
\title{Propulsion Systems for a Starshade Rendezvous Mission with the Habitable World Observatory}

\author[a,*]{Mario Damiano}
\author[a]{Rhonda Morgan}
\author[a]{Michael Turmon}
\author[b]{Dmitry Savransky}
\author[a]{Armen Tokadjian}

\affil[a]{Jet Propulsion Laboratory, California Institute of Technology, Pasadena, CA 91109, USA}
\affil[b]{Department of Astronomy, Cornell University, 122 Sciences Drive, Ithaca, NY 14853, USA}

\cftpagenumbersoff{figure}
\cftpagenumbersoff{table} 
\begin{document} 
\maketitle

\begin{abstract}
    We assess how propulsion and operational constraints shape the exo-Earth characterization performance of a starshade rendezvous architecture for a Habitable Worlds Observatory (HWO)–like telescope at Sun–Earth L2. Using end-to-end Monte Carlo mission simulations with EXOSIMS, we model two representative starshade point designs: a 60 m occulter optimized for visible-band (VIS) spectroscopy (500–1000 nm) and a 35 m occulter optimized for near-ultraviolet (NUV) characterization (250–550 nm). For the 60 m case, we compare three electric-propulsion options spanning Hall and gridded-ion technologies (AEPS Hall, NEXT ion, and a Starlink-class Hall system) and quantify yield sensitivity to onboard propellant capacity, retargeting trajectory “burn-portion” (accelerate and decelerate fraction of slew), and the presence or absence of on-orbit refueling. In fuel-limited scenarios (1000–3000 kg initial propellant), the higher-Isp NEXT system delivers the largest characterization yield, providing 9 to 13\% more characterizations than AEPS and 13 to 16\% more than the Starlink-class option by enabling more retargeting slews per kilogram of propellant. When refueling is allowed, cumulative yields converge across thruster choices (about 120 cumulative characterizations over a 10-year campaign), while the propulsion choice primarily trades against the number of refueling events. For the 35 m near-UV starshade (modeled with AEPS), yields are substantially less fuel-sensitive and are instead dominated by mission time and geometric accessibility; tightening the minimum solar keepout (scarf) angle from 40$^{\circ}$ toward $\geq$ 50$^{\circ}$ barely impacts yield, from 50$^{\circ}$ to 65$^{\circ}$ modestly impacts yield, and $\geq$ 65$^{\circ}$reduces yield rapidly, with severe degradation near 80$^{\circ}$. These results identify regimes where starshade performance is fundamentally fuel-limited versus time/keepout-limited and provide quantitative guidance for jointly optimizing propulsion selection, tank size, refueling policy, and keepout constraints in HWO starshade architectures.
\end{abstract}


{\noindent \footnotesize\textbf{*} Corresponding author,  \linkable{mario.damiano@jpl.nasa.gov} }

\begin{spacing}{2}   

\section{Introduction}
\label{sec:intro}  

Direct imaging of exoplanets aims to spatially separate a faint planetary signal from the overwhelmingly brighter glare of the host star in order to obtain reflected- or emitted-light spectra that reveal atmospheric composition. Unlike indirect techniques (e.g., radial velocities or transits), direct imaging uniquely enables spectroscopy at moderate resolving power across visible and ultraviolet (UV) bands, opening access to key features such as \ce{O2} and \ce{O3} as well as Rayleigh slopes and surface/ocean glint diagnostics for temperate, potentially habitable worlds. The ultimate science goal is the robust detection and atmospheric characterization of Earth-size planets in the habitable zones of nearby Sun-like stars.

The 2020 Decadal Survey in Astronomy and Astrophysics (\emph{Astro2020}) recommended the development of a large infrared/optical/ultraviolet (IR/O/UV) observatory with \(\sim\)6\,m aperture optimized for high-contrast imaging and spectroscopy of Earth-like exoplanets while also enabling transformative general astrophysics \citep{NASEM2021Astro2020}. In response, the Habitable Worlds Observatory (HWO) concept has been initiated: a mission that consolidates heritage from LUVOIR and HabEx to deliver the wavefront quality, stability, and starlight suppression required for robust exo-Earth detection and atmosphere characterization.

Both external occulters (“starshades”) and internal coronagraphs are capable of suppressing starlight to directly image a nearby exoplanet.  Starshades suppress starlight before it enters the telescope by formation-flying a petal-shaped screen tens of thousands of kilometers from the observatory, thereby relaxing telescope wavefront stability requirements, providing high throughput, and offering intrinsically broadband suppression \citep{Cash2006}. Recent HWO studies have examined starshade point designs tailored to complementary wavelength ranges, including a \(\sim\)60\,m diameter starshade optimized for the visible (VIS) band (e.g., \(\sim\)500–1000\,nm) and a \(\sim\)35\,m diameter starshade optimized for the near-ultraviolet (NUV) to access ozone absorption spectral features  \citep{Shaklan2024,Shaklan2023UV}. These options illustrate how starshades could enhance HWO’s biosignature reach while retaining a coronagraph for discovery and orbit determination.

The compelling advantages of a starshade come with mission-level systems considerations. In particular, starshade operations are constrained by (i) \emph{retargeting} slews between stars and \emph{station-keeping} to maintain the formation flying transverse alignment, both of which consume propellant; (ii) onboard propellant capacity (tank size) and specific impulse; and (iii) geometric and thermal constraints such as a minimum solar keepout angle (a Sun-avoidance constraint that restricts when specific targets are observable). Retargeting typically dominates fuel use and introduces cadence overheads that couple directly to science yield \citep{Stark2016,Flinois2021}. Minimizing solar keepout angle can expand sky access and mitigate seasonal gaps, but may increase stray light backgrounds and thermal or pointing risks; overall, the yield sensitivity to keepout must therefore be quantified within realistic scheduling and propulsion models.

The propulsion and operations trades considered here are rooted in earlier starshade yield, mission-design, and technology studies \citep{Glassman2007,Benson2009,Gaudi2020HabEx,RomeroWolf2021a,RomeroWolf2021b}. Starshade yield studies have long identified retargeting fuel, slew cadence, and target visibility as first-order mission resources, with optimal performance occurring between purely fuel-limited and exposure-time-limited regimes \citep{Stark2016}. Earlier New Worlds Observer studies also identified solar electric propulsion as enabling for starshade retargeting at Sun--Earth L2 and compared candidate electric-propulsion systems for a 50\,m starshade architecture, selecting NEXT because it provided high science return with efficient xenon use, manageable system complexity, and extended-mission margin \citep{Glassman2007,Benson2009}. HabEx and Roman Space Telescope Starshade Rendezvous studies developed mission concepts in which external occulters provide broadband characterization capability and require coordinated scheduling, formation flying, and retargeting operations \citep{Gaudi2020HabEx,RomeroWolf2021a,RomeroWolf2021b}. The broader starshade literature, including the JATIS starshade special issue, provides useful background on starshade science, engineering, technology maturity, and programmatics \citep{Arenberg2021}. Parallel starshade technology studies have examined optical performance, background calibration, and formation-flying sensing requirements \citep{Hu2020,Hu2021,Bottom2020}. The present work extends this heritage by comparing current electric-propulsion options within a common HWO-like Monte Carlo yield framework and by explicitly separating fuel-limited, refueling-limited, and keepout-limited operating regimes.

In this paper we quantify these trade spaces for two representative HWO-relevant starshade designs: a 60\,m starshade for visible wavelengths and a 35\,m starshade for the near UV. For the 60\,m starshade operating in visible wavelengths, we compared (a) three different thrusters to understand how science yield is affected; (b) the dependence of science yield on fuel consumption rate and initial propellant load (onboard fuel quantity or tank size); and (c) the role of on-orbit refueling versus a no-refuel scenario. For the 35\,m starshade operating in the NUV, we explored (a) the trade space between fuel consumption and onboard fuel; and (b) the impact of the minimum solar keepout angle on the achievable detection and characterization yield. We place particular emphasis on the interplay between fuel-limited and exposure-time-limited regimes and on mission scheduler behavior when both fuel and keepout constraints are active.

Building on the previously established result that higher $I_{sp}$ reduces the propellant cost of starshade retargeting maneuvers and can increase yield in fuel-limited mission regimes \citep{soto2019fuel}, we present an HWO-specific numerical assessment of current propulsion options, tank sizes, burn-portion choices, refueling assumptions, and solar-keepout geometry under a common set of mission-simulation assumptions for the VIS and NUV point designs. The resulting yield estimates are not intended to be quantitatively representative of a finalized HWO architecture, since they depend on simplified and still-evolving assumptions regarding mission design and observing strategy. Instead, they provide a consistent basis for comparative assessment across propulsion and mission-design choices, and for identifying transitions between regimes in which retargeting performance is primarily fuel-limited and those in which finite mission duration, target availability, slew cadence, or solar keepout become the dominant constraints.

To conduct this study we use the \emph{Exoplanet Open-Source Imaging Mission Simulator} (\texttt{EXOSIMS}), an extensible, modular framework for end-to-end simulation of direct-imaging missions \citep{Savransky2016}. \texttt{EXOSIMS} assembles mission realizations from interchangeable submodules that describe target catalogs, astrophysical populations, observatory orbits, keepout geometry, instrument performance, integration-time calculators, schedulers, and—in the case of starshades—propulsion/fuel models and formation-flying constraints. The framework has been applied extensively to yield analyses and scheduler optimization for space-based direct imaging, and it natively supports the accounting needed to track fuel usage during 
station-keeping and retargeting \citep[e.g.,][]{savransky2010,soto2017starshadespie,soto2018optimal,soto2019fuel}. 
In our application, \texttt{EXOSIMS} is configured to represent HWO-class telescope parameters together with the above 60\,m (visible) and 35\,m (near UV) starshade design points, and to sample a range of fuel loads, burn rates, tank capacities, and solar keepout angles.

\paragraph{Scope and organization.} Section~\ref{sec:method} describes the mission and instrument assumptions, the \texttt{EXOSIMS} configuration (including propulsion and keepout models), and the experiment design for the parameter sweeps. Section~\ref{sec:results} presents yield and fuel-consumption results across the explored trade spaces and contrasts refueling versus no-refueling scenarios for both starshade sizes. Section~\ref{sec:discussion} discusses implications for HWO architecture, scheduler design, and technology priorities (e.g., propulsion $I_{sp}$, tank mass fraction, and keepout-driven stray light control). Section~\ref{sec:conclusion} summarizes the principal findings and outlines applications to future HWO pre-phase A studies.

\section{Method}\label{sec:method}

\subsection{Telescope (HWO-like Observatory)}
\label{sec:telescope}

We model the observatory as an HWO-class, diffraction-limited 6\,m IR/O/UV space telescope operating at Sun--Earth L2 to provide a thermally stable, low-background environment with stringent line-of-sight and wavefront stability \citep{NASEM2021Astro2020}. The science metric reported throughout this paper is starshade characterization yield. 
Unless otherwise specified, ``yield'' denotes the cumulative number of successful starshade characterization observations, including revisits, averaged over a Monte Carlo ensemble of simulated planetary universes. Each universe is generated using the same fixed catalog of real target stars, but with independently drawn planetary populations based on the adopted occurrence-rate distributions.
When we discuss unique characterized planets or revisit observations separately, we state that distinction explicitly. In the BLIND-search cases, the simulation includes the discovery and orbit-determination coronagraph-observations needed to identify targets before starshade spectroscopy. In the OMNI (omniscient) cases, the schedule consists of characterization observations of already known planets. The prior knowledge of planet existence and orbital phase could be assumed from precursor extreme-precision radial-velocity surveys, astrometry, multi-epoch direct imaging, or an earlier HWO coronagraph campaign. For Earth analogs around nearby FGK stars, the EPRV Working Group identified sub-\(10\,\mathrm{cm\,s^{-1}}\) precision with multi-year stability as the relevant measurement scale for detecting and measuring masses and orbits \citep{Crass2021}. In both BLIND and OMNI case, the reported yield remains the number of successful starshade characterizations.

\noindent\textit{\textbf{Visible IFS for starshade characterization.}}
The visible instrument is a lenslet-fed integral-field spectrograph (IFS) spanning 500--1000\,nm at resolving power \(R_s \approx 140\) with a detector pixel scale of \(0.0193''\)~pix\(^{-1}\) at 750\,nm. For a 6\,m aperture, the diffraction scale at 750\,nm is \(\lambda/D \simeq 25.8\)~mas (first Airy minimum \(1.22\,\lambda/D \simeq 31.5\)~mas), which yields \(\sim1.3\)--\(1.6\)~pixels per \(\lambda/D\) at band center. The lenslet sampling is set to two lenslets per \(\lambda/D\), meeting Nyquist criteria for lenslet IFS designs \citep[e.g.,][]{Gong2017}. The end-to-end optical throughput is taken as 0.28 (representative HabEx/HWO starshade-IFS average over 500--1000\,nm), and the detector quantum efficiency is \(\mathrm{QE}=0.90\), implying a photoelectron conversion of \(\sim0.25\). This configuration is consistent with state-of-the-art EMCCD performance in space applications \citep{Bush2025} (detailed parameters are listed in Table~\ref{tab:vis_tech_par}). With starshade suppression, noise is dominated by photon statistics from planet signal and zodiacal/exozodiacal backgrounds \citep{Hu2020}. 


\noindent\textit{\textbf{Near-UV camera for ozone-sensitive characterization.}}
To access ozone and Rayleigh-scattering diagnostics in the near-UV, we include a grism/filter-based NUV spectrograph (or slitless spectroscopy) optimized near 250--550\,nm with low resolving power \(R_s \approx 7\). The camera is modeled after a delta-doped EMCCD201 implementation (HabEx heritage, detailed parameters listed in Table~\ref{tab:nuv_tech_par}).

At 310\,nm, \(\lambda/D \simeq 10.7\)~mas for a 6\,m aperture; the \(4.4\)~mas~pix\(^{-1}\) sampling provides \(\sim2.4\)~pix per \(\lambda/D\) (and \(\gtrsim 2\)~pix per \(\lambda/D\) across 250--550\,nm), ensuring modest super-Nyquist sampling of the diffraction core. Detector noise remains sub-dominant for the long integrations used here. These assumptions are consistent with HabEx/HWO detector and coatings trades for near-UV performance \citep{Gaudi2020HabEx}.

\medskip

\subsection{Starshade}
\label{sec:starshade}

\begin{figure}[htb]
    \centering
    \includegraphics[width=0.95\linewidth]{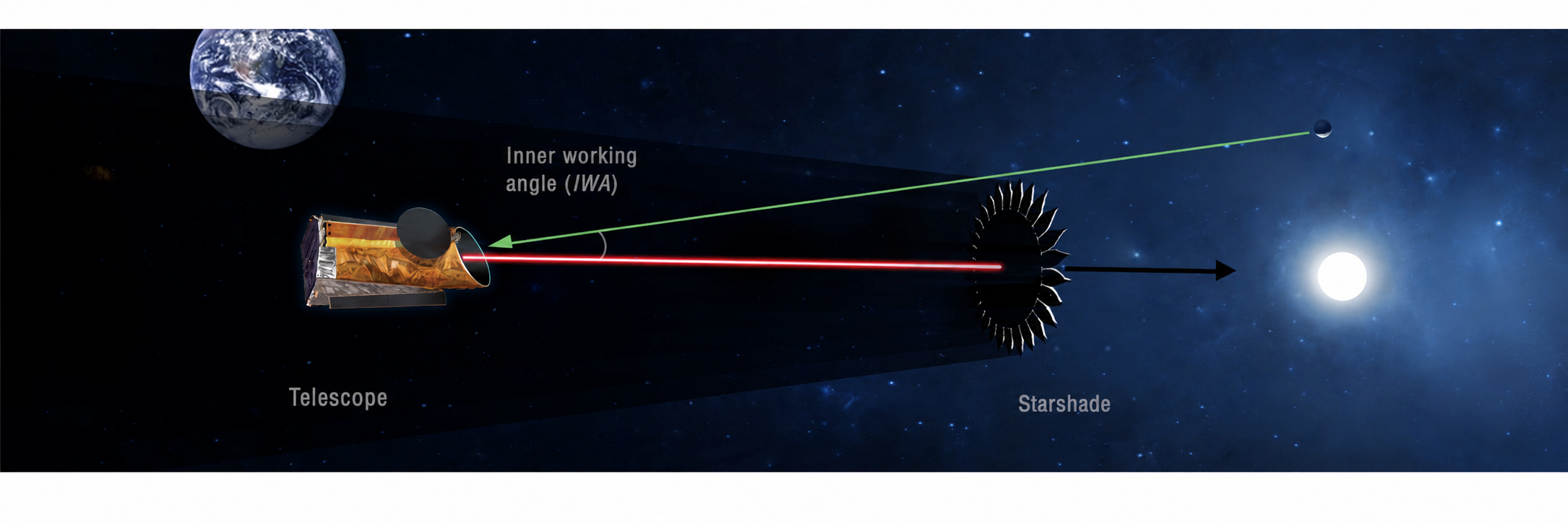}
    \caption{The telescope flying in formation with the starshade. The telescope can detect and characterize exoplanets viewable near or beyond the angular radius of the starshade as seen from the telescope, which defines the inner working angle (IWA). Figure adopted from the Habex final report \cite{Gaudi2020HabEx}.}
    \label{fig:starshade_formation}
\end{figure}

We consider two external occulter designs representative of recent HWO-Starshade point studies: a 60\,m-diameter starshade optimized for the visible band (500--1000\,nm) and a 35\,m-diameter starshade optimized for the near-UV (250--500\,nm). Both designs employ petal-shaped apodizers operating at Fresnel numbers typical of high-suppression starshades and are formation-flown at fixed telescope–occulter separations chosen to set the inner working angle (IWA) at band center (see Figure \ref{fig:starshade_formation}) \citep{Cash2006,Shaklan2024,Shaklan2023UV}.

\noindent\textit{\textbf{60\,m visible starshade.}}
The 60\,m design is tuned to a central wavelength of 750\,nm with fractional bandwidth \(\mathrm{BW}=0.67\) (500–1000\,nm). The occulter–telescope separation is fixed at 95{,}200\,km, yielding an IWA of \(0.035''\) at band center. We adopt an occulter optical throughput of 0.95 (exclusive of the telescope/instrument), and a fractional visit overhead of 0.33 to capture target acquisition, fine alignment, and detector/instrument setup. The starshade point-spread ``core'' is sampled internally at \(0.195''\)~pix\(^{-1}\) for convolution with the IFS sampling.
In this band, the visible IFS of Section~\ref{sec:telescope} is used for characterization integrations.

\noindent\textit{\textbf{35\,m near-UV starshade.}}
The 35\,m design targets a central wavelength of 418\,nm with \(\mathrm{BW}=0.67\) (280–550\,nm) and an occulter–telescope separation of 55{,}500\,km, producing an IWA of \(0.020''\) at band center. We adopt occulter optical throughput of unity (conservative upper bound; coating and scatter losses are captured in the calibration files), and the same 0.33 fractional visit overhead. Internal sampling of the starshade core is \(0.466''\)~pix\(^{-1}\).
Characterization integrations in this band use the NUV camera of Section~\ref{sec:telescope}, enabling access to UV-sensitive biomarkers (e.g., \ce{O3}) and short-wavelength continuum slopes.

\noindent\textit{\textbf{Operational notes common to both designs.}}
For both occulters, the keepout angles used in scheduling are \(\epsilon_{\odot}\in[40^\circ,83^\circ]\), with Earth and Moon keepouts of \([1^\circ,180^\circ]\). The fractional observing overhead \(\mathrm{ohTime}=0.33\) day (about 8\,hr) is applied per characterization visit. In \texttt{EXOSIMS}, the occulter optical states above are combined with the telescope/instrument throughputs of Section~\ref{sec:telescope} to compute end-to-end detection and integration-time budgets, while formation-flying constraints and keepout windows control target availability and cadence \citep{Savransky2016,Keithly2020}.

Table~\ref{tab:main_design_params} lists the principal architecture parameters used in the propulsion calculations, including the starshade dry masses. These dry masses are the non-propellant masses used by \texttt{EXOSIMS} when converting installed thrust into acceleration and when updating the wet mass as propellant is expended. The appendix tables provide the corresponding detector and instrument parameters.

\begin{table}[htbp]
    \centering
    \small
    \caption{Main starshade architecture parameters used in the mission simulations.}
    \label{tab:main_design_params}
    \begin{tabular}{lcc}
        \hline
        Parameter & 60\,m VIS starshade & 35\,m NUV starshade \\
        \hline
        Starshade dry mass & 5600 kg & 2500 kg \\
        Starshade diameter & 60 m & 35 m \\
        Telescope--starshade separation & 95{,}200 km & 55{,}500 km \\
        Central wavelength & 750 nm & 418 nm \\
        Bandpass & 500--1000 nm & 280--550 nm \\
        Mission duration modeled & 10 yr & 5 yr \\
        Propulsion options & AEPS, NEXT, Starlink-class & AEPS \\
        Installed thrusters & 3 AEPS / 8 NEXT / 12 Starlink & 1 AEPS \\
        \hline
    \end{tabular}
\end{table}

\begin{figure}[htb]
    \centering
    \includegraphics[width=0.65\linewidth]{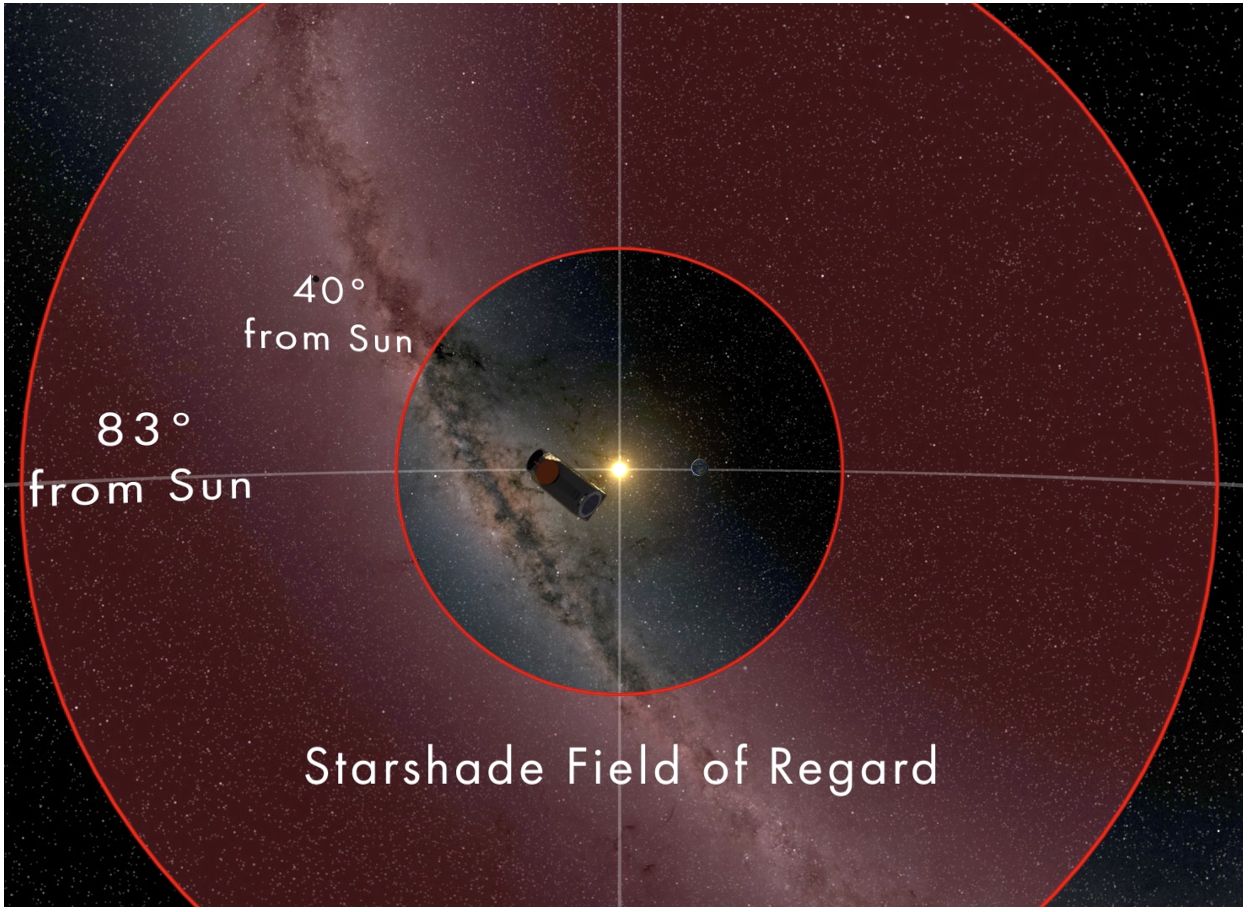}
    \caption{Field of regard for the starshade--telescope system, showing minimum and maximum solar keepout angles.}
    \label{fig:field_of_regard}
\end{figure}

Figure~\ref{fig:field_of_regard} defines the field-of-regard geometry used in the scheduler. The solar elongation \(\epsilon_{\odot}\) is the angle between the telescope boresight and the Sun direction. A scarf angle larger than the baseline \(40^\circ\) minimum keepout raises the minimum allowed solar elongation and narrows the accessible field of regard; \(90^\circ\) is normal to the Sun and \(180^\circ\) is anti-Sun.

These two designs bracket complementary wavelength ranges for HWO-class characterization and permit direct comparison of yield sensitivity to IWA, separation distance, and keepout geometry under otherwise similar scheduling and detector assumptions \citep{Cash2006,Shaklan2024,Shaklan2023UV}.

\subsection{Starshade thrusters}
\label{sec:thrusters}

Efficient re-targeting and precise station-keeping for a formation-flying starshade are propulsion–limited tasks: the available thrust \(T\) sets the characteristic acceleration \(a = T/m\) for a starshade of mass \(m\), and the specific impulse \(I_{\rm sp}\) sets the propellant mass flow \(\dot m = T/(I_{\rm sp} g_0)\) and, via the rocket equation, the propellant cost of a required \(\Delta v\) \citep{sutton2001rocket}. To quantify the coupling between propulsion and science yield, we evaluated three flight-relevant electric propulsion options spanning Hall and ion technologies and distinct propellants (xenon and argon). The values reported in Table \ref{tab:thrusters} are those used in our \texttt{EXOSIMS} experiments; they are representative of published performance envelopes for the respective thruster families (Hall effect and gridded-ion) and provide a consistent basis for scheduling and fuel accounting \citep[e.g.,][]{Choueiri2009,Brophy2002GridErosion}.


The use of high-\(I_{\rm sp}\) electric propulsion for starshade retargeting is motivated by both starshade-specific trade studies and flight and technology heritage from solar-electric propulsion missions and development programs. In the New Worlds Observer study \citep{Benson2009}, the authors compared NEXT, NSTAR, XIPS-25, and BPT-4000 options for a starshade mission requiring \(\sim7\)--10\,km\,s\(^{-1}\) total velocity change and found that NEXT provided the most favorable combination of xenon efficiency, science return, lifetime margin, and system simplicity. More broadly, the NSTAR ion engine demonstrated long-duration xenon ion propulsion on Deep Space 1 and Dawn, while NEXT and AEPS/HERMeS represent later high-power gridded-ion and Hall-effect systems relevant to large-spacecraft retargeting and deep-space operations \citep{Rayman1999,Sovey2001,Brophy2009Dawn,Herman2018}.

The Starlink-class case is used as an illustrative argon Hall-effect propulsion proxy with publicly discussed performance characteristics, not as a detailed model of a proprietary flight unit. It is included to bracket the consequences of a lower-\(I_{\rm sp}\), lower-per-thruster-thrust Hall option relative to AEPS and NEXT under otherwise identical scheduling assumptions.
Unless stated otherwise, re-targeting burns assume symmetric firings that allow all installed units to contribute without violating torque/pointing constraints, yielding a maximum net thrust \(T_{\rm max} = N_{\rm thr}\,T\). Station-keeping is modeled with duty-cycled thrust at the single-thruster level.

\begin{table}[htbp]
    \centering
    \small
    \caption{ Technical properties of the three propulsion systems compared in this work. }
    \label{tab:thrusters}
    \begin{tabular}{lccc}
        \hline
        Option & Propellant & \(I_{\rm sp}\) (s) & \(T\) (mN) \\
        \hline
        AEPS Hall & Xe & 2900 & 600 \\
        NEXT ion  & Xe & 4190 & 236 \\
        Starlink HET & Ar & 2500 & 170 \\
        \hline
    \end{tabular}
\end{table}


\noindent While the NEXT option offers the highest \(I_{\rm sp}\) (and thus the lowest propellant consumption per unit \(\Delta v\)), both Hall options provide larger installed thrust per kilogram of system mass in typical implementations, shortening low-thrust slews at fixed starshade mass and target geometry. This trade between time-to-retarget and fuel-per-\(\Delta v\) is central to the yield outcomes in Section~\ref{sec:results}.
The number of thrusters installed on the starshade is constrained by the solar-array collecting area on the central disk. For the 60\,m starshade, the assumed installations are three AEPS Hall thrusters for 1800\,mN total thrust, eight NEXT thrusters for 1888\,mN total thrust, and twelve Starlink-class Hall thrusters for 2040\,mN total thrust. For the 35\,m starshade, we use a single AEPS Hall thruster with 600\,mN total thrust.

\noindent\textit{\textbf{Use in EXOSIMS.}}
In the propulsion submodel, retargeting between two targets separated by an angular distance \(\Delta\psi\) at telescope--occulter range \(z\) is approximated as a planar lateral translation of length \(s \approx z\,\Delta\psi\), where \(\Delta\psi\) is measured in radians and denotes the target-to-target separation on the sky. We use \(\Delta\psi\), rather than \(\theta\), to distinguish this slew angle from the solar-elongation and scarf-angle quantities used in the keepout model.

For a symmetric burn--coast--burn slew, the acceleration magnitude is \(a=T_{\rm eff}/m\), where \(T_{\rm eff}\) is the effective thrust available for the retargeting maneuver after accounting for the number of operating thrusters, thrust-vector geometry, pointing constraints, and duty cycle, and \(m\) is the instantaneous wet mass. The total slew duration is \(t_{\rm slew}\), and \(\beta\) denotes the burn portion, defined as the fraction of the total slew time spent thrusting during the acceleration and deceleration arcs. The resulting one-dimensional constant-acceleration relation is
\begin{equation}
s = \frac{a\,t_{\rm slew}^{2}}{4}\,\beta(2-\beta),
\qquad
t_{\rm slew} =
\left[\frac{4s}{a\,\beta(2-\beta)}\right]^{1/2}.
\label{eq:slew_time}
\end{equation}
Here \(0<\beta\le1\); \(\beta=1\) is the no-coast triangular profile, while smaller \(\beta\) values introduce a coast interval and lengthen the slew at fixed \(s\), \(m\), and \(T_{\rm eff}\).

For each finite burn segment \(j\), the propellant mass is computed from the thrust duration and the instantaneous mass flow,
\begin{equation}
\Delta m_j = \frac{T_{{\rm eff},j}\Delta t_j}{I_{\rm sp}g_0},
\label{eq:mass_flow}
\end{equation}
with the spacecraft wet mass updated after each burn. Equivalently, for an impulsive equivalent \(\Delta v_j\), the rocket equation gives
\begin{equation}
\Delta m_j =
m_j\left[1-\exp\left(-\frac{\Delta v_j}{I_{\rm sp}g_0}\right)\right],
\label{eq:rocket_segment}
\end{equation}
where \(m_j\) is the wet mass before segment \(j\). Thus, the fuel use is not assumed to be the same for every slew; it varies with target separation, burn time, effective thrust, and the decreasing wet mass.

A useful limiting case follows if every characterization visit requires the same effective \(\Delta v\), the dry mass is \(S\), the initial fuel mass is \(F\), and all other scheduling constraints are ignored. Summing Eq.~\ref{eq:rocket_segment} over \(N\) identical visits and solving for $N$ gives
\begin{equation}
N_{\rm lim} =
\frac{I_{\rm sp}g_0}{\Delta v}
\ln\left(1+\frac{F}{S}\right).
\label{eq:analytic_visits}
\end{equation}
This analytic expression shows the expected linear dependence on \(I_{\rm sp}\) for fixed \(S\), \(F\), and \(\Delta v\). The purpose of the \texttt{EXOSIMS} simulations is to move beyond this limiting case by allowing \(\Delta v_j\), target availability, keepout windows, exposure times, burn portion, mission duration, and scheduler choices to vary across the mission.

The fuel model includes both retargeting slews and station-keeping during aligned observations. Retargeting dominates the propellant use in the cases studied here, but station-keeping is retained because differential lateral acceleration between the telescope and starshade sets the control effort needed to maintain alignment during observations \citep{Kulik2021}. More broadly, recent cislunar formation-flying studies emphasize that long-duration multi-spacecraft architectures require explicit treatment of background dynamics, relative-motion control, and missed-maneuver robustness; however, the station-keeping costs used here are tied to the starshade-specific Sun--Earth L2 geometry and EXOSIMS fuel model rather than to a generic cislunar-orbit survey \citep{Foss2025Cislunar}.

The refueling-enabled cases are modeled as idealized upper bounds on science yield. In these simulations, a refueling event restores the available propellant to the specified tank capacity but does not include overhead time. 
This choice isolates the propulsion-yield coupling from a specific servicing architecture. Fuel servicing architecture development and studies do not yet define the overhead time or transfer efficiency for a deep-space xenon refueling architecture for starshades at Sun--Earth L2, though significant progress has been made in in-space fuel transfer demonstrations and aspects such as autonomous rendezvous, robotic servicing, and fluid-transfer operations \citep{mulvaney2025space,Tomlinson2022RRM,Henshaw2022Grappling, herman2022development}. Autonomous in-space fuel transfer was demonstrated in 2007 by DARPA's Orbital Express architecture \citep{Shoemaker2003OrbitalExpress} which introduced a standardized interface for fuel transfer.  NASA's Robotic Refueling Mission 3 (RRM3) aboard the International Space Station demonstrated in 2018-2019 long-term cryogenic lossless storage of liquid methane and robotic manipulation of a cryogenic servicing tool and visual inspection tool \citep{McGuire2019RRM3}. SpaceX demonstrated cryogenic transfer of a significant amount of liquid oxygen between two tanks in 2024 \citep{Hauser2025Cryogenic}. Future capability is heading toward commercial refueling services with multiple future demonstrations planned \citep{mulvaney2025space}. As such, we have chosen to analyze a zero-downtime limiting case; finite servicing downtime would reduce the absolute yields but would not change the basic conclusion that refueling primarily trades propulsion choice against the number of servicing events.

\subsection{Universe}
\label{sec:universe}

We generate the astrophysical scene with \texttt{EXOSIMS} synthesizing planets around a catalog of real nearby stars, and propagating the synthetic planets to the epochs of observation, and calculating the signal and noise photometry observed by the instrument model \citep{Savransky2016}. The objective in this work is to isolate propulsion/scheduling impacts on \emph{characterization} yield for Earth analogs; accordingly, we employ an Earths-only population prescription. For the simulations in the VIS broadband, we followed the definition of an exo-Earth first used in the HabEx and LUVOIR concept studies \citep{Gaudi2020HabEx, luvoir2019, morgan2019sdetfinal} and continued use in HWO analysis \citep{morgan2026}: the semi-major axis $a$ is in the conservative habitable zone \cite{kopparapu2013, kopparapu2018} between 0.95 and 1.67 AU and the planetary radius is between $0.8/\sqrt{a}$  and 1.4 $R_{\oplus}$. In the NUV simulations, we redefined the planetary radius in line with the Fulton Gap definition \cite{Fulton2017}, pushing the radius upper bound to 1.75 $R_{\oplus}$. Because the NUV simulations adopt a larger planet-radius upper bound than the VIS simulations, direct VIS/NUV yield comparisons are not controlled solely for starshade diameter, wavelength range, and propulsion architecture. We therefore interpret the VIS/NUV comparison as an architecture-level comparison under the adopted population definitions, rather than as an isolated wavelength-only effect.

\noindent\textit{\textbf{Target stars.}}
The target list comprises cataloged nearby stars with measured astrometric and photometric properties (distance, luminosity class, broadband magnitudes, etc). We use the ExoCat-1 catalog \citep{turnbull2015exocat} that was originally derived from Hipparcos stars in the nearest 30 pc. These stellar parameters feed the instrument and starshade models (Sections~\ref{sec:telescope}--\ref{sec:starshade}) and set the geometry for habitable-zone scaling, phase functions, and illumination. All field-of-regard observation windows and solar keepout constraints are computed from the real sky coordinates of each star within \texttt{EXOSIMS}.

\noindent\textit{\textbf{Planet occurrence prior (Earths-only).}}
Planet semi-major axis and radius are drawn from the joint-probability occurrence rate density histogram of  \citep[e.g.,][]{DulzPlavchan2020}. Stars with exo-Earths most commonly have a single terrestrial planet consistent with an ``exo-Earth'' definition (Earth-like size/insolation, circular-orbit baseline with modest eccentricity scatter), a small portion have two, and very rarely a star has three. Semi-major axes and habitable zone (HZ) are scaled by the square root of the host star luminosity. Orbital orientations are isotropic. The radii drawn from the joint occurrence rate distribution are converted to mass using the Chen and Kipping 2016 mass–radius relationship \cite{chen2017probabilistic}. The reflected-light signal uses a gray geometric albedo of 0.2 for rocky planets and a Lambertian phase function, ensuring self-consistency across bands.

\noindent\textit{\textbf{Monte Carlo realizations.}}
We obtain a Monte Carlo ensemble of $N_\mathrm{sim}=100$ mission instances 
by initializing statistically-independent universes simulations.
In each simulation, we generate a distinct cloud of Earth analogs over the same fixed stellar catalog by resampling the planet-occurrence distributions and orbital elements. Mission scheduling (including starshade slews and keepouts) is then executed on each realization with identical telescope/instrument/propulsion settings.
We emphasize that the 
chosen target visits are not random draws; they are directed by 
a dynamically responsive scheduling algorithm \cite{savransky2010, morgan2021}. The specific planet configurations and orbital phases differ across 
ensemble members (Monte Carlo generated planets), and the scheduler selects observations from the available target stars subject to integration-time, keepout, fuel, and slew-time constraints.

Each mission in the ensemble has a final characterization yield, and 
the resulting ensemble mean yield is the basis for the results we report. 
We also track its Monte Carlo sampling uncertainty via the sample standard deviation, 
$\sigma_\textrm{yield}$. 
The standard error of the mean yield then 
equals \(\sigma_{\rm yield}/\sqrt{N_\textrm{sim}}\), 
which proves sufficient to resolve the multi-parameter trends reported in Section~\ref{sec:results} while keeping computational expense tractable.
Indeed, the Monte Carlo approach allows us to find the 
mean and corresponding sampling error of all simulated 
quantities, including fuel usage and slew times.

\section{Results} \label{sec:results}
\subsection{VIS wavelength band}
Starting with the VIS band (500--1000\,nm), we tested how the three propulsion systems affect characterization yield in fuel-limited cases. We considered three initial propellant loads: 1000, 2000, and 3000\,kg.

\begin{figure}[h]
    \centering
    \includegraphics[width=0.95\linewidth]{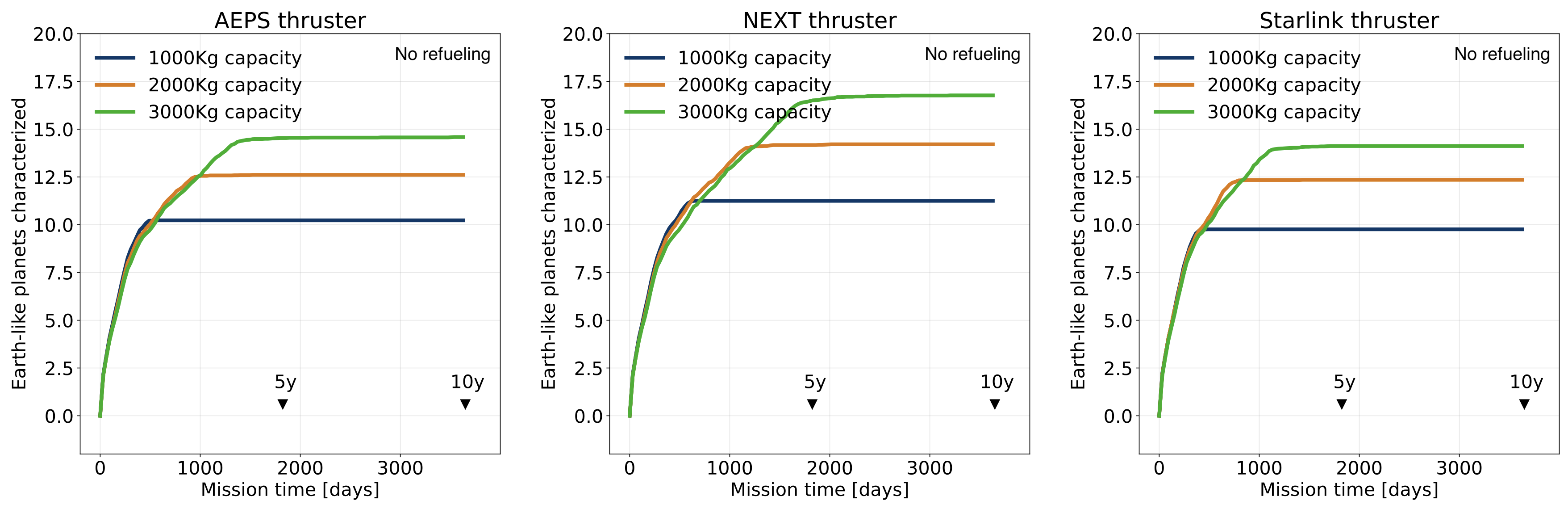}
    \caption{Characterization yield as a function of mission time for three onboard fuel budgets (1000, 2000, and 3000\,kg) for the three different propulsion systems. Refueling was turned off for these examples, therefore these represent fuel-limited cases.} The simulations assume identical telescope and keepout parameters; only thruster performance and propellant consumption differ.
    \label{fig:thrusters_fixed_tank}
\end{figure}

As shown in Fig.~\ref{fig:thrusters_fixed_tank}, the characterization yield increases with mission time for all configurations, rising steadily as the starshade performs characterization observations and eventually flattening once the available fuel is exhausted. Larger tank sizes extend the operational window, enabling the starshade to conduct more observations and thus characterize a greater number of planets.  

Across the three propulsion options and for all initial fuel capacities (1000, 2000, and 3000\,kg), the NEXT thruster provides the best overall performance, yielding roughly 9–13\% more characterizations than AEPS and 13–16\% more than the Starlink-class system. This advantage derives from NEXT’s higher specific impulse: its efficient propellant usage slows fuel consumption and allows the mission to approach five years of characterization time with a 3000\,kg tank. 

\begin{figure}[h]
    \centering
    \includegraphics[width=0.75\linewidth]{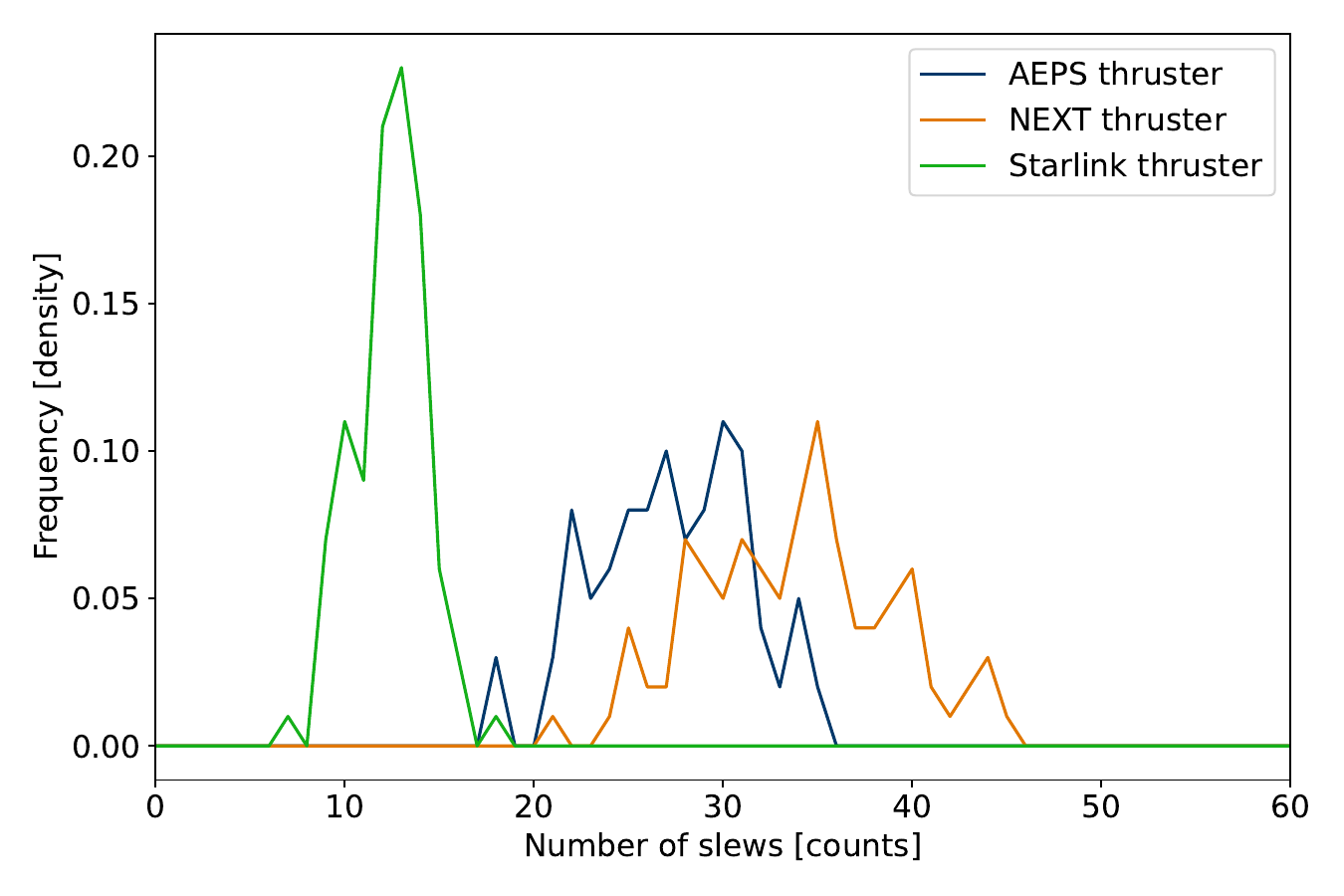}
    \caption{Monte Carlo distribution of slew counts for the three thrusters for the 2000 kg tank and no refueling case. The NEXT thruster achieves the highest number of slews with the available fuel, resulting in more characterization observations.}
    \label{fig:slews_fixed_tank}
\end{figure}

The Monte Carlo distribution of slew counts provides a complementary view of this behavior. In Fig.~\ref{fig:slews_fixed_tank}, the NEXT distribution is shifted toward higher values relative to AEPS and the Starlink-class system, indicating that across the 100 simulated universes it performs more slews on average. This increased slew capacity produces a larger number of characterization observations in the fixed-tank cases.

\begin{figure}[h]
    \centering
    \includegraphics[width=0.95\linewidth]{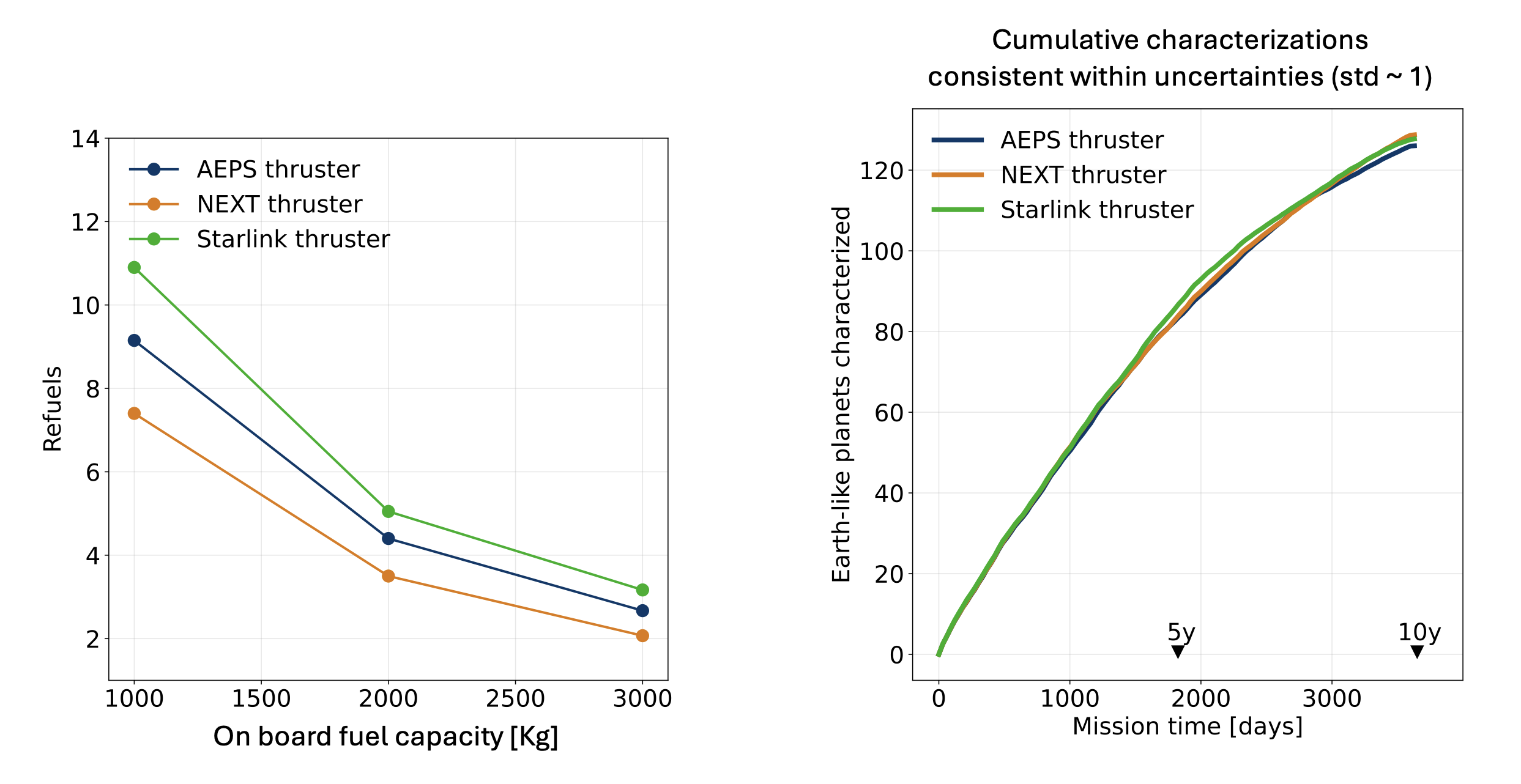}
    \caption{\textbf{Left panel:} number of refuels required to complete the 10-year mission as a function of tank size. The NEXT thruster remains the most efficient option. \textbf{Right panel:} characterization yield as a function of mission time, where the three thrusters exhibit largely comparable performance.}
    \label{fig:thrusters_vis_refuel}
\end{figure}

In the idealized zero-overhead refueling cases, the three thrusters converge to nearly identical characterization yields, reaching roughly 120 cumulative characterizations over the 10-year mission regardless of tank size. Figure~\ref{fig:thrusters_vis_refuel} (right panel) shows this convergence directly: once propellant can be replenished without time penalty, differences in fuel efficiency and slew cadence no longer drive meaningful yield separation. The left panel instead shows the number of refueling events required as a function of tank capacity. The Starlink-class case requires more frequent refueling because its lower \(I_{\rm sp}\) consumes propellant more rapidly during retargeting despite its shorter retargeting times. AEPS exhibits intermediate behavior, while NEXT remains the most propellant-efficient option, requiring the fewest refueling events across all configurations.

\begin{figure}[h]
\centering
    \includegraphics[width=0.95\linewidth]{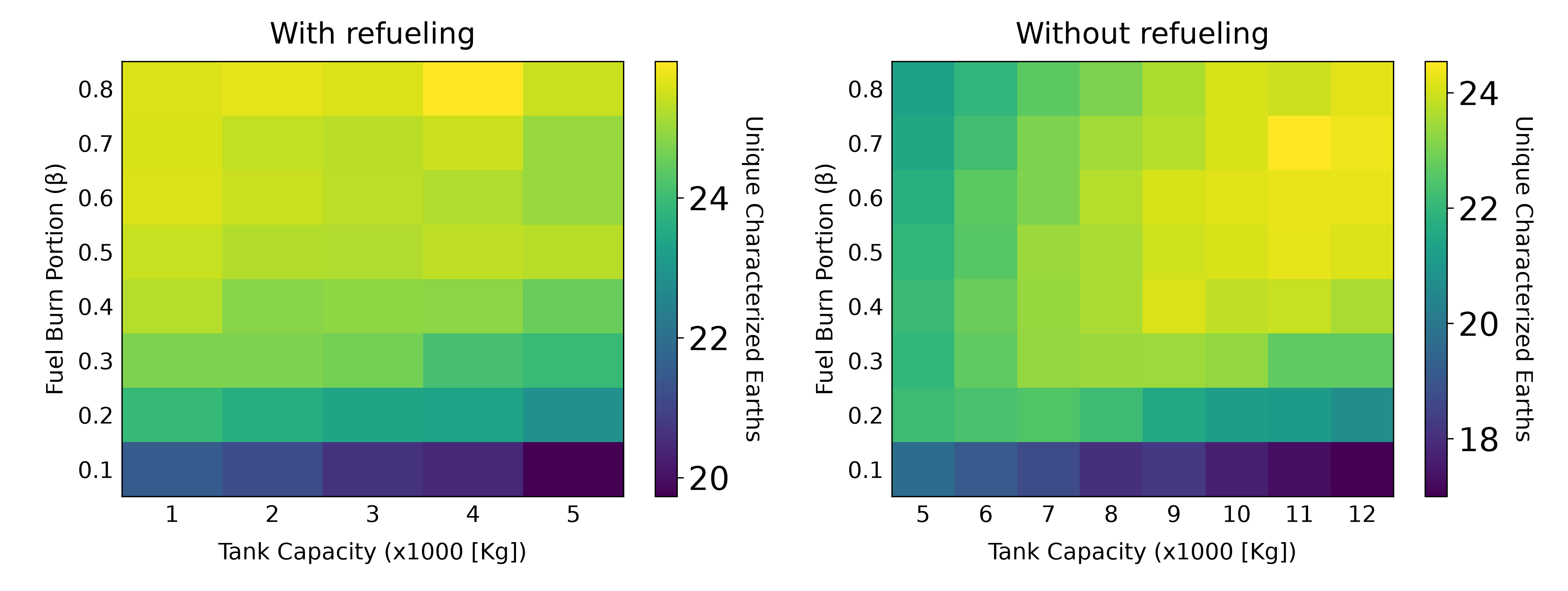}
    \caption{Characterization yield as a function of tank size and burn-portion ratio, shown with refueling enabled (left panel) and disabled (right panel). Note - the characterization yield here is defined as unique characterized Earths, i.e., the number of Earth-like planets characterized at least once.}
    \label{fig:next_db_tank}
\end{figure}

After examining these results, we focused on the NEXT thruster and explored how characterization yield varies with tank size and burn-portion ratio \(\beta\) (Eq.~\ref{eq:slew_time}). The purpose of this sweep is not to show that larger \(I_{\rm sp}\) improves fuel-limited performance, which follows from Eq.~\ref{eq:analytic_visits}, but to identify where the numerical schedule departs from the simple fuel-limited expectation because wet mass, slew time, finite mission duration, and target availability also matter. Figure~\ref{fig:next_db_tank} summarizes the optimization outcomes for both the refueling and no-refueling cases. With refueling enabled, the concern of exhausting propellant is removed, and the yield peaks at high burn-portion ratios. The maximum occurs at a 4000\,kg tank rather than 5000\,kg because the additional wet mass slows the average slew time and slightly reduces overall cadence.

When refueling is disabled, larger tank sizes and higher burn-portion ratios generally produce higher yields until the schedule approaches the time-limited regime. The maximum number of characterizations in this sweep is achieved with an 11{,}000\,kg tank and \(\beta=0.7\), with comparable yields obtained at even larger tank capacities and \(\beta=0.8\), or with slightly smaller tanks paired with lower burn portions (i.e., tank of 9,000 and 10,000kg with $\beta=0.6$).

\subsection{NUV wavelength band}

For the NUV simulations, we model the 35\,m starshade with AEPS propulsion and examine how characterization yield varies with tank size and burn-portion ratio. We do not include refueling in these cases because the lower-mass 35\,m design can complete the modeled 5-year mission with substantially lower propellant sensitivity than the 60\,m design. We evaluate both BLIND-search and OMNI scenarios.

\begin{figure}[h]
\centering
    \includegraphics[width=0.95\linewidth]{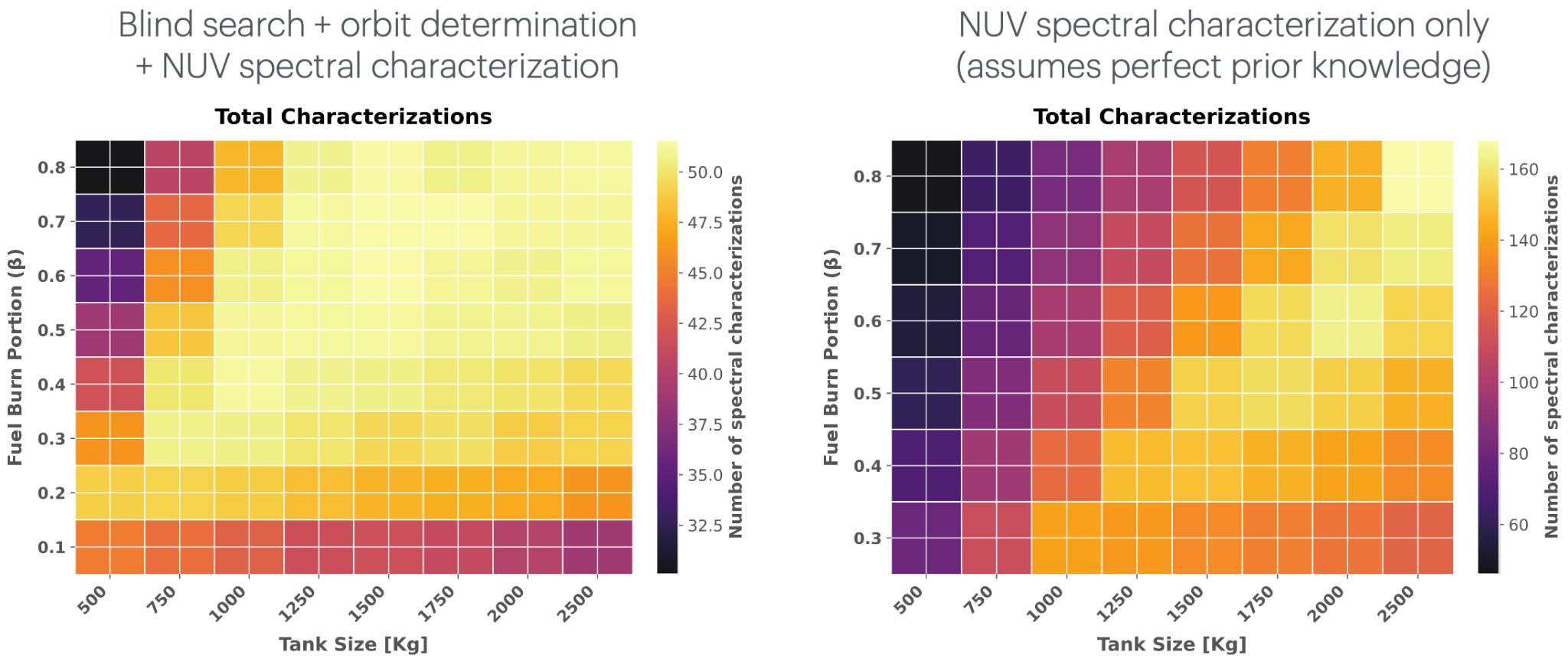}
    \caption{Characterization yield as a function of tank size and burn portion for the 35\,m starshade operating in the NUV band. \textbf{Left panel:} BLIND-search scenario, in which the telescope must first discover planets and constrain their orbits before starshade characterization. \textbf{Right panel:} OMNI scenario, in which the planets are assumed to be known and only characterization observations are performed.}
    \label{fig:next_db_tank_nuv}
\end{figure}

Figure~\ref{fig:next_db_tank_nuv} summarizes the findings. As in the VIS case, characterization yield initially increases with larger tank sizes and higher burn portions. In the BLIND-search scenario (left panel), the time available for characterization is reduced because the telescope must first detect planets and constrain their orbit. This produces a broader range of tank-size and burn-portion combinations with near-maximal characterization yield. The maximum yield occurs for a tank size of \(1500\pm250\,\mathrm{kg}\) and \(\beta=0.7\pm0.1\).

In the OMNI scenario, prior knowledge of the planet population allows the scheduler to perform characterization observations without an initial discovery phase. Two solutions emerge as particularly favorable in this sweep: (1) a 2000\,kg tank with \(\beta=0.6\), and (2) a 2500\,kg tank with \(\beta=0.8\).

Figure~\ref{fig:next_db_tank} refers to the 60\,m starshade operating for a 10-year OMNI mission, whereas the 35\,m starshade shown in Fig.~\ref{fig:next_db_tank_nuv} is simulated for a 5-year mission. The 35\,m NUV case yields more characterizations in both BLIND-search and OMNI scenarios (while considering for both scenarios the unique characterization yield as in Figure~\ref{fig:next_db_tank}) because its smaller dry mass and shorter telescope--starshade separation reduce slew times and propellant sensitivity. This comparison should be interpreted as an architecture-level result, not as a wavelength-only effect.

\subsubsection{Field of regard exploration}

In all preceding simulations, the starshade was allowed to operate over the baseline field of regard, \(\epsilon_{\odot}=40^\circ\)--\,\(83^\circ\), defined in Fig.~\ref{fig:field_of_regard}. For the 35\,m starshade, we also examined how the characterization yield responds when the minimum allowed scarf angle is increased, thereby narrowing the field of regard.

\begin{figure}[h]
    \centering
\textbf{}    \includegraphics[width=0.95\linewidth]{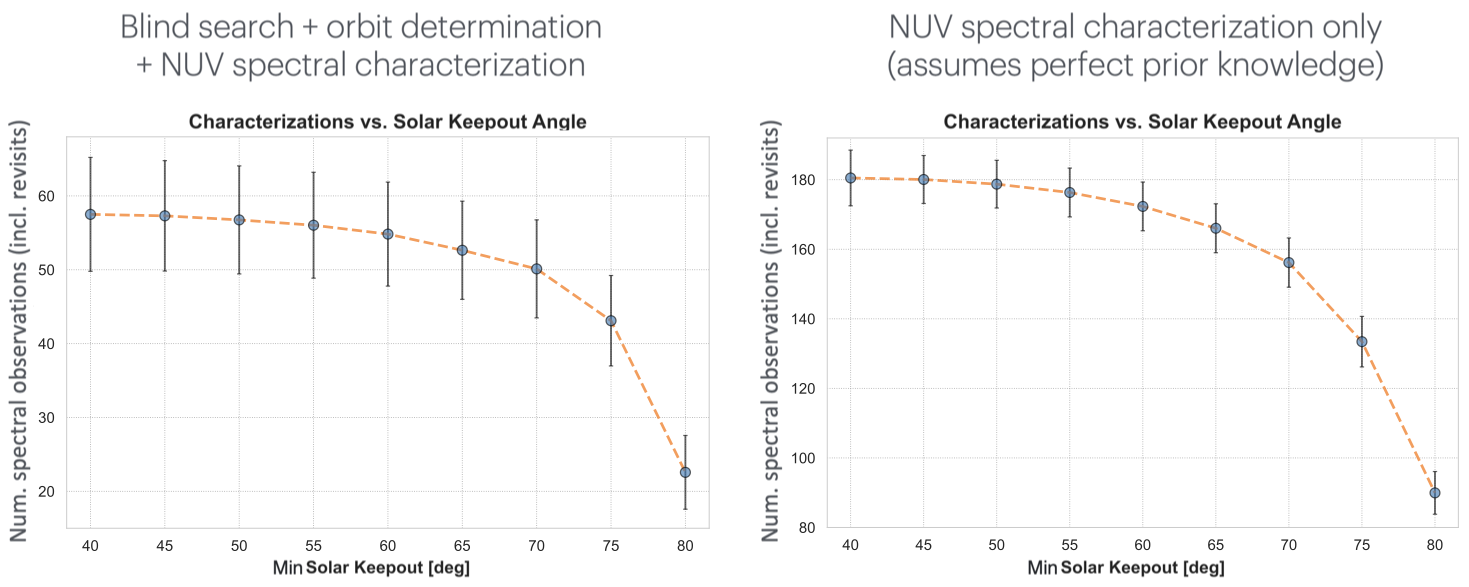}
    \caption{Characterization yield as a function of minimum scarf angle for the 35\,m starshade.
    Error bars show $\sigma_\textrm{yield}$, the 1-$\sigma$ scatter of the ensemble. Errors of the mean yield are lower by a factor of $\sqrt{N_\textrm{sim}}=10$.}
    \label{fig:scarf_angle}
\end{figure}

Figure~\ref{fig:scarf_angle} shows how characterization yield varies as the minimum scarf angle is increased while the maximum solar elongation is fixed at \(83^\circ\). The yield decreases by approximately 1\% as the minimum scarf angle rises from \(40^\circ\) to \(50^\circ\), then declines more rapidly beyond \(60^\circ\). At \(80^\circ\), the allowable field of regard is narrow enough that few targets remain observable, producing a pronounced yield loss. The same qualitative trend is observed in both BLIND-search and OMNI scenarios.

The scarf-angle study considers only the geometric solar-keepout constraint. Thus, the dependence shown in Fig.~\ref{fig:scarf_angle} is entirely a consequence of target accessibility and scheduling geometry, rather than any assumed change in starshade optical performance.

\subsection{35\,m starshade point-design study}
To illustrate the behavior of a representative 35\,m starshade architecture, we examine one point design from Fig.~\ref{fig:next_db_tank_nuv}: a 2500\,kg tank and \(\beta=0.8\) burn-portion ratio without refueling in the OMNI scenario. Because the OMNI scenario assumes that all planets in the simulated universe are already known, the scheduler can prioritize repeated retargeting and characterization rather than discovery.

\begin{figure}[h]
    \centering
    \includegraphics[width=0.80\linewidth]{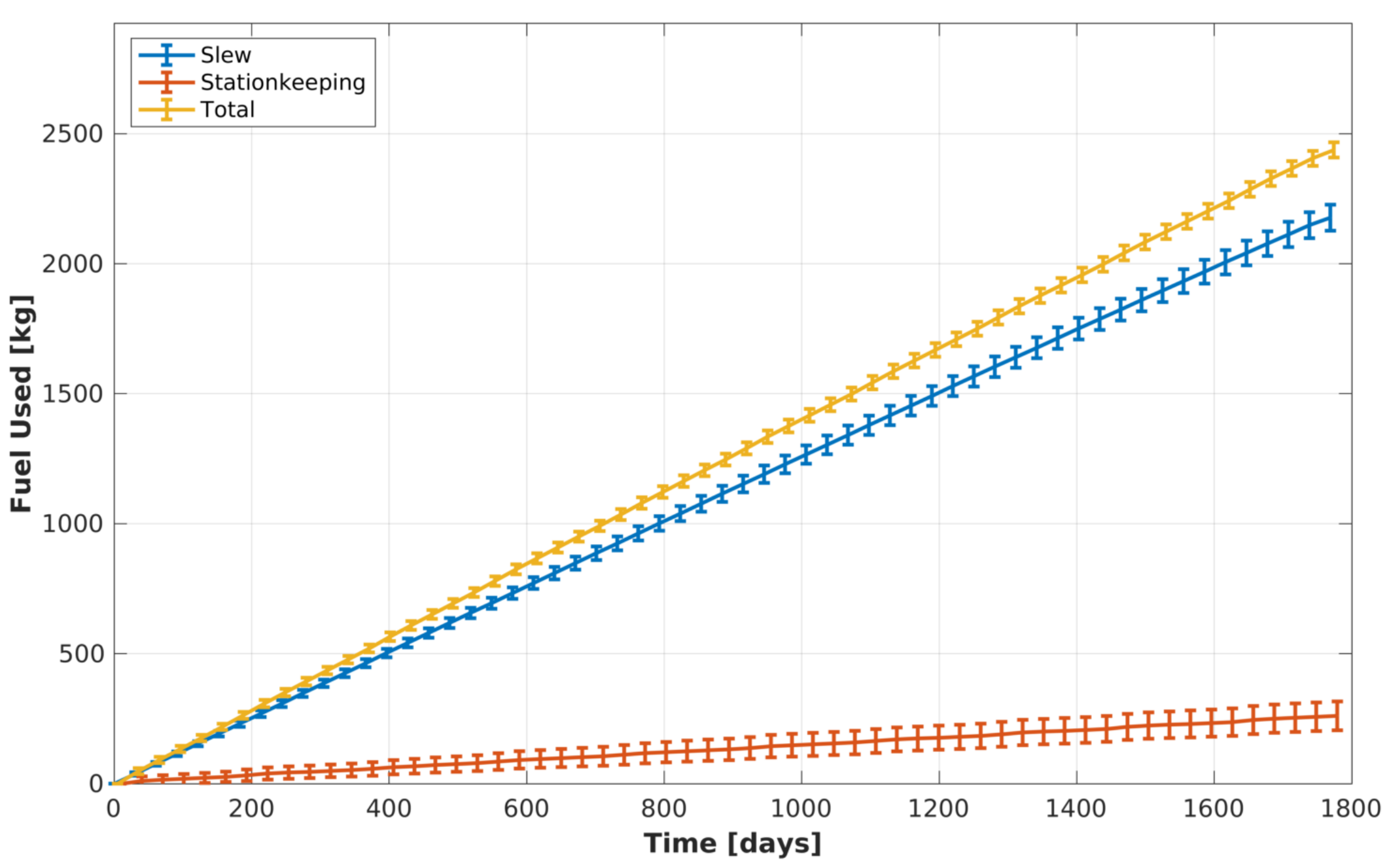}
    \caption{Fuel consumption as a function of mission time for the 35\,m starshade under the OMNI scenario. Cumulative fuel consumption for slews and stationkeeping is binned at a monthly granularity and averaged across the 100-member Monte Carlo ensemble (see sec.~\ref{sec:universe}). Error bars represent the 1-$\sigma$ scatter of the ensemble. Contraction of total fuel error bars at mission's end is due to the tendency of near-complete fuel exhaustion at that time.}
    \label{fig:35m_ss_fuel}
\end{figure}

Figure~\ref{fig:35m_ss_fuel} shows fuel consumption as a function of mission time. The propellant budget is dominated by retargeting slews, while station-keeping during aligned characterization observations contributes a smaller share. In this point design, the 2500\,kg initial propellant load is consumed over the 5-year mission.

\begin{figure}[h]
    \centering
    \includegraphics[width=0.80\linewidth]{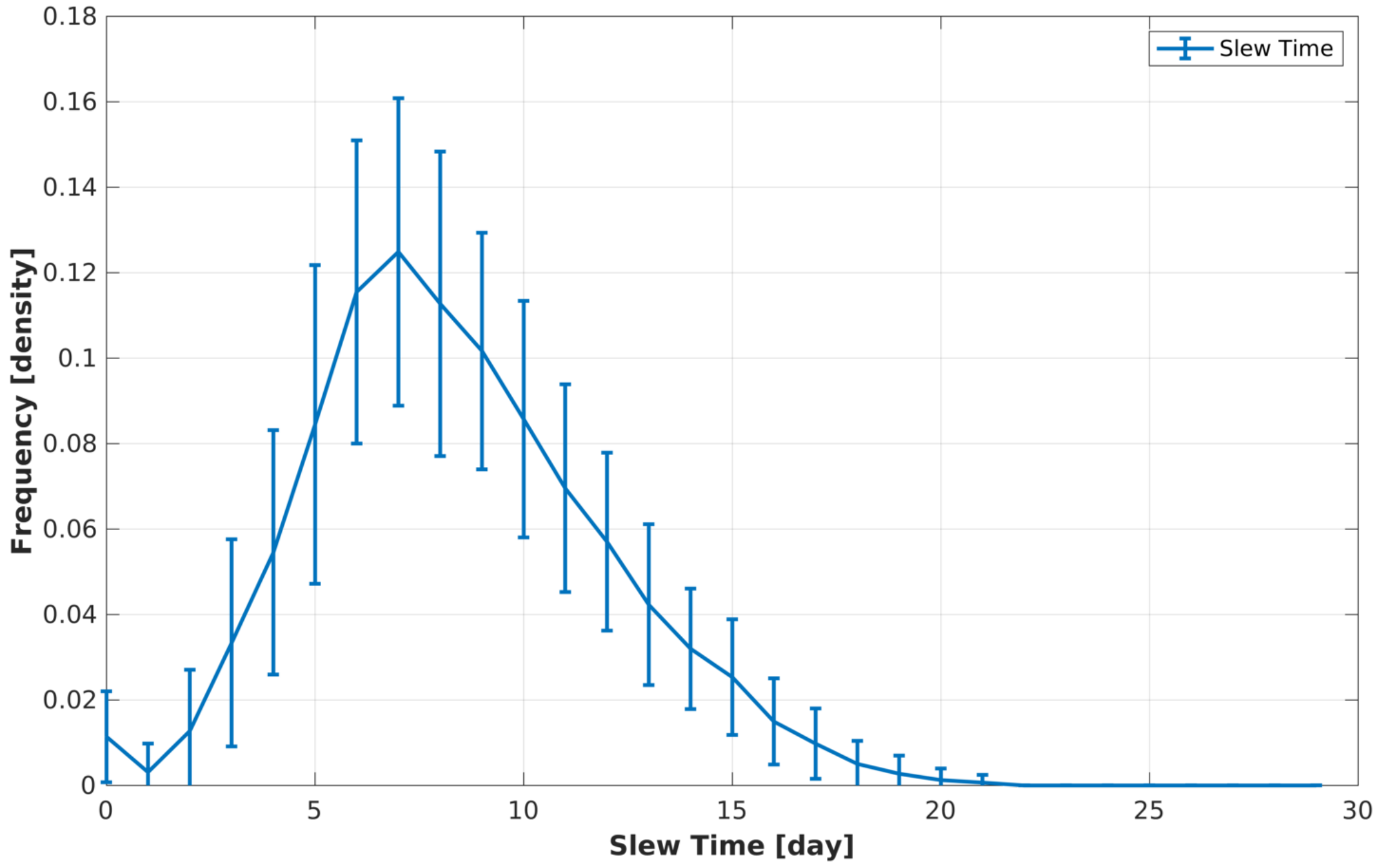}
    \caption{Slew-duration histogram across the Monte Carlo simulations for the 35\,m starshade. Error bars represent the 1-$\sigma$ Monte Carlo scatter of the ensemble.}
    \label{fig:35m_ss_slew}
\end{figure}

For the 35\,m starshade, the combination of dry mass, available thrust, and telescope--starshade separation results in relatively short retargeting times. Figure~\ref{fig:35m_ss_slew} shows the distribution of slew durations across the Monte Carlo simulations. The distribution peaks near 7 days, with most slews spanning approximately 1--20 days. During starshade repositioning, the observatory could potentially conduct other science activities, including coronagraphic planet searches or general astrophysics observations.

\begin{figure}[h]
    \centering
    \includegraphics[width=0.80\linewidth]{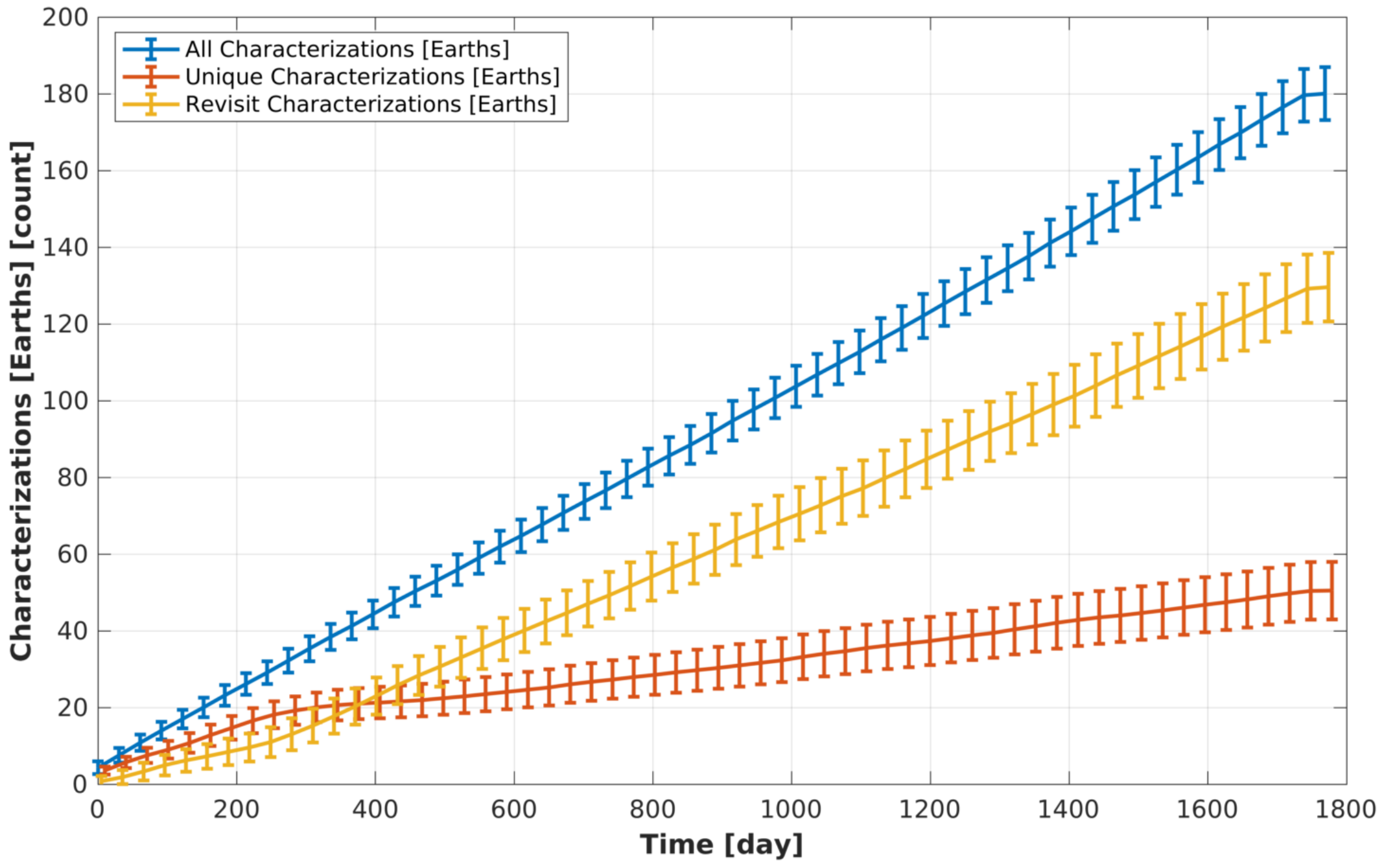}
    \caption{Number of characterized Earth-like planets as a function of mission time for the 35\,m starshade. Error bars represent the 1-$\sigma$ scatter of the ensemble.}
    \label{fig:35m_ss_char}
\end{figure}

For this point design, the 35\,m starshade performs \(\sim\)50 unique Earth-analog characterizations and \(\sim\)130 revisit characterizations, for a total yield of \(\sim\)180 characterization observations over five years (Fig.~\ref{fig:35m_ss_char}).

\section{Discussion} \label{sec:discussion}

The characterization yields obtained across our simulation campaigns show how propulsion choice, starshade size, and solar keepout jointly shape the mission performance of a formation-flying observatory such as HWO. Across all fixed-tank VIS simulations, the three thrusters follow an ordering driven by the trade between installed thrust and propellant efficiency: Starlink-class propulsion offers the fastest slews but consumes propellant at the highest rate; NEXT provides the slowest slews but maximizes xenon efficiency; and AEPS sits between these two regimes. As a result, when fuel is limited, NEXT outperforms the other thrusters because its higher specific impulse enables more retargetings per kilogram of xenon. AEPS supplies intermediate performance, while Starlink-class thrusters achieve competitive cadence only when tank sizes are large or refueling is available. Once idealized zero-overhead refueling is enabled, these distinctions largely collapse, leading all three thrusters to converge to similar cumulative yields while differing mainly in the number of required refueling events.

Within this broader landscape, optimization of tank size and burn-portion ratio reveals another important contrast between the VIS and NUV starshade designs. The heavier 60\,m VIS starshade benefits most from large tanks and high burn-fraction trajectories, as the longer slew distances and slower accelerations make propellant a primary driver of mission yield. In contrast, the lower-mass 35\,m NUV starshade operates in a regime where slews are faster, propellant consumption is lower, and mission time--rather than fuel--becomes the dominant constraint. This produces a broader plateau of near-optimal solutions, particularly in the BLIND-search scenario where the telescope must first discover planets before spectral characterization with the starshade.

The dependence of yield on \(I_{\rm sp}\) is therefore not linear. Higher \(I_{\rm sp}\) increases the number of retargetings available per kilogram of propellant, but the marginal benefit decreases once the mission becomes limited by slew duration, exposure time, target availability, or keepout geometry. This behavior is consistent with earlier starshade scheduling analyses, which found maximum efficiency between the fuel-limited and exposure-time-limited limits \citep{Stark2016}.

The solar keepout constraint also shapes starshade scheduling by determining target accessibility: as the allowable scarf-angle range narrows, fewer targets are available for observation, with yields decreasing more rapidly once the minimum keepout angle approaches \(\sim\!60^\circ\).

For a mission like HWO, these results imply that propulsion system selection and keepout policy are not independent levers but jointly determine the feasible scheduling envelope, particularly when integrating a starshade into a multi-band characterization roadmap. High-\(I_{\rm sp}\) systems such as NEXT offer the most robust performance in fuel-limited modes, while a wide range of scarf angles and their associated solar keepout angles (40-60 degrees) preserves access to the target catalog.

A starshade rendezvous architecture requires observatory-level interfaces for formation sensing, relative navigation, and coordinated operations. These implementation details are not modeled in the present study, which focuses on science yield and mission geometry. Formation-flying, optical alignment sensing, and related operational concepts have been studied extensively in dedicated starshade technology investigations (e.g., \citep{Bottom2020}), and their detailed accommodation by HWO is therefore beyond the scope of this work. The present analysis therefore treats formation flying as an enabling mission capability rather than as an additional yield-limiting factor.

\section{Conclusions} \label{sec:conclusion}

Across this work, we used \texttt{EXOSIMS} end-to-end mission simulations to explore how propulsion choice, starshade size, and solar keepout jointly determine the performance of a starshade-enabled Habitable Worlds Observatory. By modeling both a 60\,m VIS and a 35\,m NUV occulter, and by sampling tank sizes, burn-portion ratios, and retargeting geometries, the analysis captures how the interplay among dry mass, propellant load, thrust, spectral bandpass, and target accessibility shapes achievable characterization yields in BLIND-search and OMNI scenarios. This framework isolates which aspects of the architecture are fuel-limited versus time- or geometry-limited, and how scheduling constraints emerge from the combined telescope--starshade system.

The results show a consistent ranking among propulsion systems and a clear divergence between the VIS and NUV architectures. In fuel-limited regimes, NEXT outperforms AEPS and Starlink-class systems due to its high specific impulse, enabling more retargetings per kilogram of xenon, as expected from the analytic limiting case in Eq.~\ref{eq:analytic_visits}. AEPS provides intermediate performance, while Starlink-class thrusters achieve competitive yields only when tank sizes are large or refueling is available. Once refueling is enabled, all three converge to similar cumulative yields, demonstrating that slew-rate advantages alone do not overcome propellant inefficiency. At the architecture level, the lower-mass 35\,m NUV starshade yields more characterizations across both mission modes because faster slews and reduced fuel sensitivity shift the limiting factor from propellant to mission time and target accessibility.

Overall, these results do not by themselves establish the full HWO exo-Earth science yield, but they provide quantitative guidance for the propulsion and operations trades that must be carried into future HWO starshade architecture studies. These findings highlight the importance of jointly optimizing propulsion, fuel policy, and solar keepout when integrating a starshade into HWO's multi-band roadmap. High-\(I_{\rm sp}\) systems such as NEXT provide the most robust performance in fuel-limited regimes, while typical scarf-angles preserve access to the target catalog. The 35\,m NUV starshade, in particular, appears promising under the modeled assumptions because its lower mass and shorter separation reduce slew-time and fuel penalties.

\subsection* {Code, Data, and Materials Availability}
\texttt{EXOSIMS} is an open source code available on GitHub: \href{https://github.com/dsavransky/EXOSIMS}{https://github.com/dsavransky/EXOSIMS}

\subsection* {Acknowledgments}
The authors thank Rashied Amini and Keith R. Warfield for helpful discussion and the anonymous referees who helped improving the quality of the manuscript. This research was carried out at the Jet Propulsion Laboratory, California Institute of Technology, under a contract with the National Aeronautics and Space Administration (80NM0018D0004). The High Performance Computing resources used in this investigation were provided by funding from the JPL Information and Technology Solutions Directorate.

\noindent Portions of this text were refined with the assistance of an AI language model (Microsoft Copilot), which was used for editing and phrasing. After using this tool/service, the author(s) reviewed and edited the content as needed and take(s) full responsibility for the content of the publication.

\subsection* {Disclosures}
The authors declare that there are no financial interests, commercial affiliations, or other potential conflicts of interest that could have influenced the objectivity of this research or the writing of this paper.

\newpage
\appendix

\section{VIS technical parameters}
The table below shows the parameter adopted to simulate a HWO-like telescope and Starshade operating in the VIS broadband.

\begin{table}[htbp]
    \centering
    \small
    \caption{Summary of the technical parameters used throughout the simulations for the VIS broadband.}
    \label{tab:vis_tech_par}
    \begin{tabular}{lcl}
        \hline
        \textbf{Telescope} & & \\
         & & \\
        Pupil diameter & \(6\) m & HWO-like primary \\
        Spectral band & 500--1000 nm & Visible, starshade-optimized \\
        Resolving power & \(R_s=140\) & IFS spectral sampling \\
        Pixel scale & \(0.0193\,^{\prime\prime}\,\mathrm{pix}^{-1}\) & At \(750\,\mathrm{nm}\) \\
        Pixel size & \(12\,\mu\mathrm{m}\) & Detector physical pitch \\
        Lenslet sampling & 2 per \(\lambda/D\) & Nyquist lenslet sampling \\
        Optical throughput & 0.28 & Telescope+IFS average \\
        Detector QE & 0.90 & Peak visible deep-depletion QE \\
        Read noise & 0 e\(^{-}\) & Photon-counting \\
        Dark current & \(3\times10^{-5}\) e\(^{-}\)/pix/s & \\
        CIC & \(1.3\times10^{-3}\) e\(^{-}\)/pix/frame & \\
        PC selection efficiency & 0.75 & Coincidence/threshold losses \\
         & & \\
        \textbf{Starshade} & & \\
         & & \\
        Diameter & 60 m & \\
        Bandwidth & 67\% & \\
        Starshade-telescope distance & 95 200 km & \\
        Dry mass & 5600 kg & \\
        Propulsion & Table \ref{tab:thrusters} & Explored multiple options \\
        \hline
    \end{tabular}
\end{table}

\newpage

\section{NUV technical parameters}
The table below shows the parameter adopted to simulate a HWO-like telescope and Starshade operating in the NUV broadband.

\begin{table}[htbp]
    \centering
    \small
    \caption{Summary of the technical parameters used throughout the simulations for the NUV broadband.}
    \label{tab:nuv_tech_par}
    \begin{tabular}{lcl}
        \hline
        \textbf{Telescope} & & \\
         & & \\
        Pupil diameter & \(6\) m & HWO-like primary \\
        Spectral band & 280--550 nm & Blue edge used at 280\,nm in stellar atlas fits \\
        Resolving power & \(R_s = 7\) & Grism/filters in series \\
        Pixel scale & \(4.4\times10^{-3}\,^{\prime\prime}\,\mathrm{pix}^{-1}\) & At \(310\,\mathrm{nm}\) \\
        Pixel size & \(13\,\mu\mathrm{m}\) & CCD201 pitch \\
        Lenslet sampling & 1 per \(\lambda/D\) &  \\
        Optical throughput & 0.40 & \\
        Detector QE & 0.55 & Delta-doped EMCCD201 (HabEx Fig.\,11.7-4) \\
        Read noise & 0 e\(^{-}\) & Photon-counting \\
        Dark current & \(1\times10^{-5}\) e\(^{-}\)/pix/s & \\
        CIC & 0 e\(^{-}\)/pix/frame & Long-frame PC, conservative \\
        PC selection efficiency & 0.75 & Threshold/coincidence losses \\
         & & \\
        \textbf{Starshade} & & \\
         & & \\
        Diameter & 35 m & \\
        Bandwidth & 67\% & \\
        Starshade-telescope distance & 55 500 km & \\
        Dry mass & 2500 kg & \\
        Propulsion & AEPS Hall &  \\
        Thruster unit & 1 & \\
        Thruster propellant & Xenon & \\
        Thrust [mN] & 600 & \\
        Thruster specific impulse [s] & 2900 & \\
        \hline
    \end{tabular}
\end{table}

\newpage

\bibliography{report}   

@ARTICLE{kopparapu2018,
       author = {{Kopparapu}, Ravi Kumar and {H{\'e}brard}, Eric and {Belikov}, Rus and {Batalha}, Natalie M. and {Mulders}, Gijs D. and {Stark}, Chris and {Teal}, Dillon and {Domagal-Goldman}, Shawn and {Mandell}, Avi},
        title = "{Exoplanet Classification and Yield Estimates for Direct Imaging Missions}",
      journal = {\apj},
         year = 2018,
        month = apr,
       volume = {856},
       number = {2},
          eid = {122},
        pages = {122},
          doi = {10.3847/1538-4357/aab205},
archivePrefix = {arXiv},
       eprint = {1802.09602},
 primaryClass = {astro-ph.EP}
}

@ARTICLE{kopparapu2013,
       author = {{Kopparapu}, Ravi Kumar and {Ramirez}, Ramses and {Kasting}, James F. and {Eymet}, Vincent and {Robinson}, Tyler D. and {Mahadevan}, Suvrath and {Terrien}, Ryan C. and {Domagal-Goldman}, Shawn and {Meadows}, Victoria and {Deshpande}, Rohit},
        title = "{Habitable Zones around Main-sequence Stars: New Estimates}",
      journal = {\apj},
         year = 2013,
        month = mar,
       volume = {765},
       number = {2},
          eid = {131},
        pages = {131},
          doi = {10.1088/0004-637X/765/2/131},
archivePrefix = {arXiv},
       eprint = {1301.6674},
 primaryClass = {astro-ph.EP}
}

@book{NASEM2021Astro2020,
  author    = {{National Academies of Sciences, Engineering, and Medicine}},
  title     = {Pathways to Discovery in Astronomy and Astrophysics for the 2020s},
  year      = {2021},
  publisher = {The National Academies Press},
  doi       = {10.17226/26141}
}

@article{Cash2006,
  author  = {Cash, W.},
  title   = {Detection of Earth-like planets around nearby stars using a petal-shaped occulter},
  journal = {Nature},
  year    = {2006},
  volume  = {442},
  pages   = {51--53},
  doi     = {10.1038/nature04930}
}

@inproceedings{Shaklan2024,
author = {Stuart Shaklan and Doug Lisman and Dylan McKeithen and Philip Dumont and Rhonda Morgan and Zahra Ahmed and Sara Seager},
title = {{Starshade concepts for the Habitable Worlds Observatory}},
volume = {13092},
booktitle = {Space Telescopes and Instrumentation 2024: Optical, Infrared, and Millimeter Wave},
editor = {Laura E. Coyle and Shuji Matsuura and Marshall D. Perrin},
organization = {International Society for Optics and Photonics},
publisher = {SPIE},
pages = {130925D},
year = {2024},
doi = {10.1117/12.3020798}
}

@inproceedings{Shaklan2023UV,
author = {Stuart Shaklan and Dylan McKeithen and Doug Lisman and Philip Dumont},
title = {{A starshade for NASA’s habitable worlds observatory concept: exoplanet characterization in the UV}},
volume = {12680},
booktitle = {Techniques and Instrumentation for Detection of Exoplanets XI},
editor = {Garreth J. Ruane},
organization = {International Society for Optics and Photonics},
publisher = {SPIE},
pages = {1268010},
year = {2023},
doi = {10.1117/12.2677793}
}

@article{Savransky2016,
       author = {{Savransky}, Dmitry and {Garrett}, Daniel},
        title = "{WFIRST-AFTA coronagraph science yield modeling with EXOSIMS}",
      journal = {Journal of Astronomical Telescopes, Instruments, and Systems},
         year = 2016,
        month = jan,
       volume = {2},
          eid = {011006},
        pages = {011006},
          doi = {10.1117/1.JATIS.2.1.011006},
archivePrefix = {arXiv},
       eprint = {1511.02869},
 primaryClass = {astro-ph.IM}
}

@ARTICLE{Stark2016,
       author = {{Stark}, Christopher C. and {Shaklan}, Stuart and {Lisman}, Doug and {Cady}, Eric and {Savransky}, Dmitry and {Roberge}, Aki and {Mandell}, Avi M.},
        title = "{Maximized exoEarth candidate yields for starshades}",
      journal = {Journal of Astronomical Telescopes, Instruments, and Systems},
         year = 2016,
        month = oct,
       volume = {2},
          eid = {041204},
        pages = {041204},
          doi = {10.1117/1.JATIS.2.4.041204},
archivePrefix = {arXiv},
       eprint = {1605.04915},
 primaryClass = {astro-ph.IM}
}

@article{Flinois2021,
author = {Thibault L. B. Flinois and Daniel P. Scharf and Carl R. Seubert and P. Doug Lisman},
title = {{Efficient starshade retargeting architecture using chemical propulsion}},
volume = {7},
journal = {Journal of Astronomical Telescopes, Instruments, and Systems},
number = {2},
publisher = {SPIE},
pages = {021206},
year = {2021},
doi = {10.1117/1.JATIS.7.2.021206}
}

@article{Keithly2020,
author = {Dean R. Keithly and Dmitry Savransky and Daniel Garrett and Christian Delacroix and Gabriel Soto},
title = {{Optimal scheduling of exoplanet direct imaging single-visit observations of a blind search survey}},
volume = {6},
journal = {Journal of Astronomical Telescopes, Instruments, and Systems},
number = {2},
publisher = {SPIE},
pages = {027001},
year = {2020},
doi = {10.1117/1.JATIS.6.2.027001}
}

@article{Gaudi2020HabEx,
       author = {{Gaudi}, B. Scott and {Seager}, Sara and {Mennesson}, Bertrand and {Kiessling}, Alina and {Warfield}, Keith and {Cahoy}, Kerri and {Clarke}, John T. and {Domagal-Goldman}, Shawn and {Feinberg}, Lee and {Guyon}, Olivier and {Kasdin}, Jeremy and {Mawet}, Dimitri and {Plavchan}, Peter and {Robinson}, Tyler and {Rogers}, Leslie and {Scowen}, Paul and {Somerville}, Rachel and {Stapelfeldt}, Karl and {Stark}, Christopher and {Stern}, Daniel and {Turnbull}, Margaret and {Amini}, Rashied and {Kuan}, Gary and {Martin}, Stefan and {Morgan}, Rhonda and {Redding}, David and {Stahl}, H. Philip and {Webb}, Ryan and {Alvarez-Salazar}, Oscar and {Arnold}, William L. and {Arya}, Manan and {Balasubramanian}, Bala and {Baysinger}, Mike and {Bell}, Ray and {Below}, Chris and {Benson}, Jonathan and {Blais}, Lindsey and {Booth}, Jeff and {Bourgeois}, Robert and {Bradford}, Case and {Brewer}, Alden and {Brooks}, Thomas and {Cady}, Eric and {Caldwell}, Mary and {Calvet}, Rob and {Carr}, Steven and {Chan}, Derek and {Cormarkovic}, Velibor and {Coste}, Keith and {Cox}, Charlie and {Danner}, Rolf and {Davis}, Jacqueline and {Dewell}, Larry and {Dorsett}, Lisa and {Dunn}, Daniel and {East}, Matthew and {Effinger}, Michael and {Eng}, Ron and {Freebury}, Greg and {Garcia}, Jay and {Gaskin}, Jonathan and {Greene}, Suzan and {Hennessy}, John and {Hilgemann}, Evan and {Hood}, Brad and {Holota}, Wolfgang and {Howe}, Scott and {Huang}, Pei and {Hull}, Tony and {Hunt}, Ron and {Hurd}, Kevin and {Johnson}, Sandra and {Kissil}, Andrew and {Knight}, Brent and {Kolenz}, Daniel and {Kraus}, Oliver and {Krist}, John and {Li}, Mary and {Lisman}, Doug and {Mandic}, Milan and {Mann}, John and {Marchen}, Luis and {Marrese-Reading}, Colleen and {McCready}, Jonathan and {McGown}, Jim and {Missun}, Jessica and {Miyaguchi}, Andrew and {Moore}, Bradley and {Nemati}, Bijan and {Nikzad}, Shouleh and {Nissen}, Joel and {Novicki}, Megan and {Perrine}, Todd and {Pineda}, Claudia and {Polanco}, Otto and {Putnam}, Dustin and {Qureshi}, Atif and {Richards}, Michael and {Eldorado Riggs}, A.~J. and {Rodgers}, Michael and {Rud}, Mike and {Saini}, Navtej and {Scalisi}, Dan and {Scharf}, Dan and {Schulz}, Kevin and {Serabyn}, Gene and {Sigrist}, Norbert and {Sikkia}, Glory and {Singleton}, Andrew and {Shaklan}, Stuart and {Smith}, Scott and {Southerd}, Bart and {Stahl}, Mark and {Steeves}, John and {Sturges}, Brian and {Sullivan}, Chris and {Tang}, Hao and {Taras}, Neil and {Tesch}, Jonathan and {Therrell}, Melissa and {Tseng}, Howard and {Valente}, Marty and {Van Buren}, David and {Villalvazo}, Juan and {Warwick}, Steve and {Webb}, David and {Westerhoff}, Thomas and {Wofford}, Rush and {Wu}, Gordon and {Woo}, Jahning and {Wood}, Milana and {Ziemer}, John and {Arney}, Giada and {Anderson}, Jay and {Ma{\'\i}z-Apell{\'a}niz}, Jes{\'u}s and {Bartlett}, James and {Belikov}, Ruslan and {Bendek}, Eduardo and {Cenko}, Brad and {Douglas}, Ewan and {Dulz}, Shannon and {Evans}, Chris and {Faramaz}, Virginie and {Feng}, Y. Katherina and {Ferguson}, Harry and {Follette}, Kate and {Ford}, Saavik and {Garc{\'\i}a}, Miriam and {Geha}, Marla and {Gelino}, Dawn and {G{\"o}tberg}, Ylva and {Hildebrandt}, Sergi and {Hu}, Renyu and {Jahnke}, Knud and {Kennedy}, Grant and {Kreidberg}, Laura and {Isella}, Andrea and {Lopez}, Eric and {Marchis}, Franck and {Macri}, Lucas and {Marley}, Mark and {Matzko}, William and {Mazoyer}, Johan and {McCandliss}, Stephan and {Meshkat}, Tiffany and {Mordasini}, Christoph and {Morris}, Patrick and {Nielsen}, Eric and {Newman}, Patrick and {Petigura}, Erik and {Postman}, Marc and {Reines}, Amy and {Roberge}, Aki and {Roederer}, Ian and {Ruane}, Garreth and {Schwieterman}, Edouard and {Sirbu}, Dan and {Spalding}, Christopher and {Teplitz}, Harry and {Tumlinson}, Jason and {Turner}, Neal and {Werk}, Jessica and {Wofford}, Aida and {Wyatt}, Mark and {Young}, Amber and {Zellem}, Rob},
        title = "{The Habitable Exoplanet Observatory (HabEx) Mission Concept Study Final Report}",
      journal = {arXiv e-prints},
         year = 2020,
        month = jan,
          eid = {arXiv:2001.06683},
        pages = {arXiv:2001.06683},
          doi = {10.48550/arXiv.2001.06683},
archivePrefix = {arXiv},
       eprint = {2001.06683},
 primaryClass = {astro-ph.IM}
}

@inproceedings{Gong2017,
author = {Qian Gong and Tyler Groff and Avi Mandell and Michael McElwain and Maxime Rizzo and Prabal Saxena},
title = {{Flight Integral Field Spectrograph (IFS) optical design for WFIRST coronagraphic exoplanet demonstration (Conference Presentation)}},
volume = {10400},
booktitle = {Techniques and Instrumentation for Detection of Exoplanets VIII},
editor = {Stuart Shaklan},
organization = {International Society for Optics and Photonics},
publisher = {SPIE},
pages = {104000A},
year = {2017},
doi = {10.1117/12.2273064}
}

@book{sutton2001rocket,
  title={Rocket Propulsion Elements},
  author={Sutton, G.P. and Biblarz, O.},
  isbn={9780471326427},
  lccn={00027334},
  series={A Wiley Interscience publication},
  year={2001},
  publisher={Wiley}
}

@article{Choueiri2009,
author = {Choueiri, Edgar},
year = {2009},
month = {03},
pages = {58-65},
title = {New Dawn for Electric Rockets},
volume = {300},
journal = {Scientific American},
doi = {10.1038/scientificamerican0209-58}
}

@inproceedings{Benson2009,
  author    = {Benson, Scott W. and Lo, Amy S. and Glassman, Tiffany M.},
  title     = {Solar Electric Propulsion for the New Worlds Observer Terrestrial Exoplanet Mission},
  booktitle = {45th AIAA/ASME/SAE/ASEE Joint Propulsion Conference \& Exhibit},
  year      = {2009},
  number    = {AIAA 2009-5094},
  address   = {Denver, Colorado},
  month     = aug,
  doi       = {10.2514/6.2009-5094}
}

@inproceedings{Glassman2007,
  author    = {Glassman, Tiffany M. and Lo, Amy S. and Lillie, Charles F. and Kroening, Kevin},
  title     = {New Worlds Observer Mission Design},
  booktitle = {AIAA SPACE 2007 Conference \& Exposition},
  year      = {2007},
  number    = {AIAA 2007-6221},
  address   = {Long Beach, California},
  month     = sep,
  doi       = {10.2514/6.2007-6221}
}

@inproceedings{Brophy2002GridErosion,
  author    = {Katz, Ira and Polk, James and Brophy, John and Anderson, John},
  title     = {Numerical Simulations of Ion Thruster Accelerator Grid Erosion},
  booktitle = {38th AIAA/ASME/SAE/ASEE Joint Propulsion Conference \& Exhibit},
  year      = {2002},
  number    = {AIAA 2002-4261},
  month     = jul,
  doi       = {10.2514/6.2002-4261}
}

@ARTICLE{DulzPlavchan2020,
       author = {{Dulz}, Shannon D. and {Plavchan}, Peter and {Crepp}, Justin R. and {Stark}, Christopher and {Morgan}, Rhonda and {Kane}, Stephen R. and {Newman}, Patrick and {Matzko}, William and {Mulders}, Gijs D.},
        title = "{Joint Radial Velocity and Direct Imaging Planet Yield Calculations. I. Self-consistent Planet Populations}",
      journal = {\apj},
         year = 2020,
        month = apr,
       volume = {893},
       number = {2},
          eid = {122},
        pages = {122},
          doi = {10.3847/1538-4357/ab7b73},
archivePrefix = {arXiv},
       eprint = {2003.01739},
primaryClass = {astro-ph.EP}
}

@article{Arenberg2021,
  author  = {Arenberg, Jonathan W. and
             Harness, Anthony D. and
             Jensen-Clem, Rebecca M.},
  title   = {Special Section on Starshades:
             Overview and a Dialogue},
  journal = {Journal of Astronomical Telescopes,
             Instruments, and Systems},
  volume  = {7},
  number  = {2},
  pages   = {021201},
  year    = {2021},
  doi     = {10.1117/1.JATIS.7.2.021201}
}

@article{Rayman1999,
  author  = {Rayman, Marc D. and Chadbourne, Pamela A. and Culwell, Jeffery S. and Williams, Steven N.},
  title   = {Mission design for Deep Space 1: A low-thrust technology validation mission},
  journal = {Acta Astronautica},
  year    = {1999},
  volume  = {45},
  number  = {4--9},
  pages   = {381--388},
  doi     = {10.1016/S0094-5765(99)00157-5}
}

@article{Sovey2001,
  author  = {Sovey, James S. and Rawlin, Vincent K. and Patterson, Michael J.},
  title   = {Ion Propulsion Development Projects in {U.S.}: Space Electric Rocket Test I to Deep Space 1},
  journal = {Journal of Propulsion and Power},
  year    = {2001},
  volume  = {17},
  number  = {3},
  pages   = {517--526},
  doi     = {10.2514/2.5806}
}

@article{Brophy2009Dawn,
author  = {Brophy, John R. and Garner, Charles E. and Mikes, Steven C.},
title   = {{Dawn} Ion Propulsion System: Initial Checkout After Launch},
journal = {Journal of Propulsion and Power},
year    = {2009},
volume  = {25},
number  = {6},
pages   = {1189--1202},
doi     = {10.2514/1.40480}
}

@techreport{Herman2018,
  author      = {Herman, Daniel A. and Tofil, Todd A. and Santiago, Walter and Kamhawi, Hani and Polk, James E. and Snyder, John S. and Hofer, Richard R. and Picha, Frank Q. and Jackson, Jerry and Allen, May},
  title       = {Overview of the Development and Mission Application of the Advanced Electric Propulsion System ({AEPS})},
  institution = {NASA Glenn Research Center},
  year        = {2018},
  number      = {NASA/TM--2018-219761}
}

@article{RomeroWolf2021a,
  author  = {Romero-Wolf, Andrew and Bryden, Geoffrey
             and Seager, Sara and Kasdin, N. Jeremy
             and Booth, Jeff and others},
  title   = {Starshade Rendezvous: Exoplanet Sensitivity
             and Observing Strategy},
  journal = {Journal of Astronomical Telescopes,
             Instruments, and Systems},
  volume  = {7},
  number  = {2},
  pages   = {021210},
  year    = {2021},
  doi     = {10.1117/1.JATIS.7.2.021210}
}

@article{RomeroWolf2021b,
  author = {Romero-Wolf, Andrew and Bryden, Geoffrey
            and Agnes, Greg and Arenberg, Jonathan W.
            and Bradford, Samuel Case and D'Amico, Simone
            and Debes, John and Greenhouse, Matt
            and Hu, Renyu and Matousek, Steve
            and Rhodes, Jason and Ziemer, John},
  title = {Starshade Rendezvous: Exoplanet Orbit
           Constraints from Multi-Epoch Direct Imaging},
  journal = {Journal of Astronomical Telescopes,
             Instruments, and Systems},
  volume = {7},
  number = {2},
  pages = {021219},
  year = {2021},
  doi = {10.1117/1.JATIS.7.2.021219}
}

@article{Hu2020,
    author = {Renyu Hu and Doug Lisman and Stuart Shaklan and Stefan Martin and Phil Willems and Kendra Short},
    title = {{Overview and reassessment of noise budget of starshade exoplanet imaging}},
    volume = {7},
    journal = {Journal of Astronomical Telescopes, Instruments, and Systems},
    number = {2},
    publisher = {SPIE},
    pages = {021205},
    year = {2021},
    doi = {10.1117/1.JATIS.7.2.021205}
}

@article{Hu2021,
  author  = {Hu, Renyu and Hildebrandt, Sergi R.
             and Damiano, Mario and Shaklan, Stuart
             and Martin, Stefan and Lisman, Doug},
  title   = {Starshade Exoplanet Data Challenge},
  journal = {Journal of Astronomical Telescopes,
             Instruments, and Systems},
  volume  = {7},
  number  = {2},
  pages   = {021216},
  year    = {2021},
  doi     = {10.1117/1.JATIS.7.2.021216}
}

@article{Bottom2020,
  author        = {{Bottom}, Michael and {Martin}, Stefan and {Cady}, Eric and {Davis}, Megan C. and {Flinois}, Thibault and {Scharf}, Dan and {Seubert}, Carl and {Zareh}, Shannon K. and {Shaklan}, Stuart},
  title         = {Starshade formation flying {I}: optical sensing},
  journal       = {arXiv e-prints},
  year          = {2020},
  eid           = {arXiv:2001.10575},
  pages         = {arXiv:2001.10575},
  doi           = {10.48550/arXiv.2001.10575},
  archivePrefix = {arXiv},
  eprint        = {2001.10575},
  primaryClass  = {astro-ph.IM}
}

@article{Kulik2021,
  author  = {Kulik, Jackson and
             Soto, Gabriel J. and
             Savransky, Dmitry},
  title   = {Minimal Differential Lateral Acceleration
             Configurations for Starshade Stationkeeping
             in Exoplanet Direct Imaging},
  journal = {Journal of Astronomical Telescopes,
             Instruments, and Systems},
  volume  = {8},
  number  = {1},
  pages   = {017003},
  year    = {2022},
  doi     = {10.1117/1.JATIS.8.1.017003}
}

@article{Foss2025Cislunar,
  author        = {{Foss}, Ethan and {Takubo}, Yuji and {D'Amico}, Simone},
  title         = {Long-Duration Station-Keeping Strategy for Cislunar Spacecraft Formations},
  journal       = {arXiv e-prints},
  year          = {2025},
  eid           = {arXiv:2507.19620},
  pages         = {arXiv:2507.19620},
  doi           = {10.48550/arXiv.2507.19620},
  archivePrefix = {arXiv},
  eprint        = {2507.19620},
  primaryClass  = {math.OC}
}

@article{Crass2021,
  author        = {{Crass}, Jonathan and {Gaudi}, B. Scott and {Leifer}, Stephanie and {Beichman}, Charles and {Bender}, Chad and {Blackwood}, Gary and {Burt}, Jennifer A. and {Callas}, John L. and {Cegla}, Heather M. and {Diddams}, Scott A. and {Dumusque}, Xavier and {Eastman}, Jason D. and {Ford}, Eric B. and {Fulton}, Benjamin and {Gibson}, Rose and {Halverson}, Samuel and {Haywood}, Raphaelle D. and {Hearty}, Fred and {Howard}, Andrew W. and {Latham}, David W. and {Lohner-Bottcher}, Johannes and {Mamajek}, Eric E. and {Mortier}, Annelies and {Newman}, Patrick and {Plavchan}, Peter and {Quirrenbach}, Andreas and {Reiners}, Ansgar and {Robertson}, Paul and {Roy}, Arpita and {Schwab}, Christian and {Seifahrt}, Andres and {Szentgyorgyi}, Andy and {Terrien}, Ryan and {Teske}, Johanna K. and {Thompson}, Samantha and {Vasisht}, Gautam},
  title         = {Extreme Precision Radial Velocity Working Group Final Report},
  journal       = {arXiv e-prints},
  year          = {2021},
  eid           = {arXiv:2107.14291},
  pages         = {arXiv:2107.14291},
  doi           = {10.48550/arXiv.2107.14291},
  archivePrefix = {arXiv},
  eprint        = {2107.14291},
  primaryClass  = {astro-ph.IM}
}

@inproceedings{Shoemaker2003OrbitalExpress,
  author    = {Shoemaker, James and Wright, Melissa},
  title     = {Orbital Express Space Operations Architecture Program},
  booktitle = {Space Systems Technology and Operations},
  volume    = {5088},
  pages     = {1--9},
  publisher = {SPIE},
  year      = {2003},
  doi       = {10.1117/12.499871}
}

@inproceedings{Tomlinson2022RRM,
  author    = {Tomlinson, Zakiya and Gallagher, William and Cassidy, Justin and Roberts, Brian and Facciol, Kristen and Easley, Joseph},
  title     = {Lessons for Future In-Space Telerobotic Servicing from Robotic Refueling Mission},
  booktitle = {2022 IEEE Aerospace Conference (AERO)},
  pages     = {1--17},
  year      = {2022},
  doi       = {10.1109/AERO53065.2022.9843499}
}

@inproceedings{McGuire2019RRM3,
    author = {McGuire, Jill M. and Neuman, Mark. A. and Ashmore, Matthew S. and Breon, Susan R. },
    title = {Cryogenic Refueling Demonstration on the Robotic Refueling Mission 3 (RRM3)},
    booktitle = {70th International Astronautical Congress (IAC)},
    pages = {588--598},
    year = {2019}
}

@article{Henshaw2022Grappling,
  author  = {Henshaw, Carl Glen and Glassner, Samantha and Naasz, Bo and Roberts, Brian},
  title   = {Grappling Spacecraft},
  journal = {Annual Review of Control, Robotics, and Autonomous Systems},
  year    = {2022},
  volume  = {5},
  pages   = {137--159},
  doi     = {10.1146/annurev-control-042920-011106}
}

@techreport{mulvaney2025space,
  title={In-Space Servicing, Assembly, and Manufacturing (ISAM) State of Play-2025 Edition},
  author={Mulvaney, John and Arney, Dale and Williams, Christina and Morel, Jose and Stockdale, Christopher and Whitlock, Christopher and Balaji, Vishruth},
  year={2025},
  institution = "NASA",
  number      = "20250008988"
  }

@misc{Hauser2025Cryogenic,
  author       = {Hauser, Daniel and Longmire, Nelson and
                  Wendt, Bruce and Pesich, Justin and
                  Sakowski, Barb and Tani, Hiroumi and
                  Blackmore, Lars},
  title        = {In-Space Cryogenic Propellant Transfer:
                  Modeling and Validation of Two-Phase Flow
                  Dynamics in Low-Gravity},
  year         = {2025},
  howpublished = {31st Space Cryogenics Workshop},
  address      = {Incline Village, NV, USA},
  month        = may,
  note         = {Conference abstract. NASA NTRS Document ID:
                  20250001056},
  url          = {https://ntrs.nasa.gov/citations/20250001056}
}

@inproceedings{herman2022development,
  title={Development and qualification status of the electric propulsion systems for the NASA PPE mission and gateway program},
  author={Herman, Daniel A and Gray, Timothy and Johnson, Ian and Hussein, Sarah and Winkelmann, Taylor},
  booktitle={37th International Electric Propulsion Conference},
  year={2022}
}

@ARTICLE{luvoir2019,
       author = {{The LUVOIR Team}},
        title = "{The LUVOIR Mission Concept Study Final Report}",
      journal = {arXiv e-prints},
         year = 2019,
        month = dec,
          eid = {arXiv:1912.06219},
        pages = {arXiv:1912.06219},
archivePrefix = {arXiv},
       eprint = {1912.06219},
 primaryClass = {astro-ph.IM},
       adsurl = {https://ui.adsabs.harvard.edu/abs/2019arXiv191206219T}
}

@article{morgan2026,
	Author = {{Morgan}, R. and {Turmon}, M. and {Damiano}, M. and {Savransky}, S. and et al.},
    Title = {{Revolutionizing design: optimizing HWO architecture for biosignature detection from NUV to NIR wavelengths}},
    Journal = {Journal of Astronomical Telescopes, Instruments, and Systems},
    Year = {2026},
    volume = 12,
    number =4
}

@misc{turnbull2015exocat,
	Author = {Turnbull, Margaret C},
	howpublished = {arXiv preprint arXiv:1510.01731},
	Title = {ExoCat-1: The Nearby Stellar Systems Catalog for Exoplanet Imaging Missions},
	Year = {2015}}

@misc{morgan2019sdetfinal,
	Author = {{Morgan}, Rhonda and {Savransky}, Dmitry and {Stark}, Chris and {Nielsen}, Eric},
	Howpublished = {\url{https://exoplanets.nasa.gov/system/internal_resources/details/original/1434_Standards_Team_Final_Report_20191007.pdf}},
    url = {https://exoplanets.nasa.gov/exep/studies/sdet/},
	Title = {The Standard Definitions and Evaluation Team Final Report: {A} common comparison of exoplanet yield},
	Year = {2019}}

@article{chen2017probabilistic,
  title={Probabilistic forecasting of the masses and radii of other worlds},
  author={Chen, Jingjing and Kipping, David},
  journal={The Astrophysical Journal},
  volume={834},
  number={1},
  pages={17},
  year={2017},
  publisher={The American Astronomical Society}
}

@article{soto2019fuel,
	Author = {{Soto}, G. and {Savransky}, D. and {Garrett}, D. and {Delacroix}, C.},
    Title = {{Parameterizing the Search Space of Starshade Fuel Costs for Optimal Observation Schedules}},
    Journal = {Journal of Guidance, Control, and Dynamics},
    Year = {2019},
    volume = 42,
    number =12,
    doi = {10.2514/1.G003747}}

@inproceedings{soto2018optimal,
	Author = {Soto, Gabriel J. and Keithly, Dean and Garrett, Daniel and Delacroix, Christian and Savransky, Dmitry},
	Booktitle = {Space Telescopes and Instrumentation 2018: Optical, Infrared, and Millimeter Wave},
	Organization = {International Society for Optics and Photonics},
	Pages = {106984M},
	Title = {Optimal starshade observation scheduling},
	Volume = {10698},
	Year = {2018}}

@inproceedings{soto2017starshadespie,
	Author = {{Soto}, Gabriel and {Sinha}, Amlan and {Savransky}, Dmitry and {Delacroix}, Christian and {Garrett}, Daniel},
	Booktitle = {Society of Photo-Optical Instrumentation Engineers (SPIE) Conference Series},
	Month = Sep,
	Pages = {104001U},
	Title = {{Starshade orbital maneuver study for WFIRST}},
	Volume = {10400},
	Year = 2017}

@ARTICLE{Bush2025,
       author = {{Bush}, Nathan and {Morrissey}, Patrick and {Hoenk}, Michael and {Kyne}, Gillian and {Basset}, Christophe and {Lamborn}, Andrew and {Letona}, Jocelyn and {Chen}, Weibo and {Aldrich}, David and {Nemati}, Bijan and {Gonzalez}, Guillermo and {Williams}, Peter and {Daigle}, Olivier and {Veilleux}, James and {Doyon}, Frederic and {Shortt}, Brian and {Mackie}, Ross},
        title = "{Assembly, Optimization and Calibration of the Roman Coronagraph Camera Systems: The Exoplanetary Systems Camera (EXCAM) and the Low Order Wavefront Sensing Camera (LOCAM)}",
      journal = {Journal of Astronomical Telescopes, Instruments, and Systems},
         year = 2025,
        month = jul,
       volume = {11},
          eid = {031503},
        pages = {031503},
          doi = {10.1117/1.JATIS.11.3.031503},
       adsurl = {https://ui.adsabs.harvard.edu/abs/2025JATIS..11c1503B}
}

@ARTICLE{Fulton2017,
       author = {{Fulton}, Benjamin J. and {Petigura}, Erik A. and {Howard}, Andrew W. and {Isaacson}, Howard and {Marcy}, Geoffrey W. and {Cargile}, Phillip A. and {Hebb}, Leslie and {Weiss}, Lauren M. and {Johnson}, John Asher and {Morton}, Timothy D. and {Sinukoff}, Evan and {Crossfield}, Ian J.~M. and {Hirsch}, Lea A.},
        title = "{The California-Kepler Survey. III. A Gap in the Radius Distribution of Small Planets}",
      journal = {\aj},
         year = 2017,
        month = sep,
       volume = {154},
       number = {3},
          eid = {109},
        pages = {109},
          doi = {10.3847/1538-3881/aa80eb},
archivePrefix = {arXiv},
       eprint = {1703.10375},
 primaryClass = {astro-ph.EP},
       adsurl = {https://ui.adsabs.harvard.edu/abs/2017AJ....154..109F}
}

@ARTICLE{savransky2010,
       author = {{Savransky}, Dmitry and {Kasdin}, N. Jeremy and {Cady}, Eric},
        title = "{Analyzing the Designs of Planet-Finding Missions}",
      journal = {\pasp},
         year = 2010,
        month = apr,
       volume = {122},
       number = {890},
        pages = {401},
          doi = {10.1086/652181},
archivePrefix = {arXiv},
       eprint = {0903.4915},
 primaryClass = {astro-ph.IM},
       adsurl = {https://ui.adsabs.harvard.edu/abs/2010PASP..122..401S}
}

@ARTICLE{morgan2021,
       author = {{Morgan}, Rhonda and {Savransky}, Dmitry and {Turmon}, Michael and {Mennesson}, Bertrand and {Dula}, Walker and {Keithly}, Dean and {Mamajek}, Eric E. and {Newman}, Patrick and {Plavchan}, Peter and {Robinson}, Tyler D. and {Roudier}, Gael and {Stark}, Chris},
        title = "{Faster Exo-Earth yield for HabEx and LUVOIR via extreme precision radial velocity prior knowledge}",
      journal = {Journal of Astronomical Telescopes, Instruments, and Systems},
         year = 2021,
        month = apr,
       volume = {7},
          eid = {021220},
        pages = {021220},
          doi = {10.1117/1.JATIS.7.2.021220},
       adsurl = {https://ui.adsabs.harvard.edu/abs/2021JATIS...7b1220M}
}
\bibliographystyle{spiejour}   

\listoffigures
\listoftables

\end{spacing}
\end{document}